\documentclass[journal=jpccck,manuscript=article]{achemso}

\usepackage{soul}
\usepackage{tabularx}
\usepackage{float}
\usepackage{caption}
\usepackage{subcaption}
\usepackage{tikz}
\usepackage{upgreek}

\usepackage{amsmath}
\usepackage{array}
\usepackage{multirow}
\usepackage{booktabs}

\SectionNumbersOn

\DeclareCaptionFormat{suboverlay}{\gdef\subcapoverlay{(\thesubfigure) #3\par}}
\DeclareCaptionStyle{suboverlay}{format=suboverlay}

\DeclareUnicodeCharacter{2212}{-}
\DeclareUnicodeCharacter{2005}{\hspace{0.25em}}

\author{Ethan Holbrook}
\affiliation{School of Materials Engineering and Birck Nanotechnology Center, Purdue University, West Lafayette, Indiana, 47907 USA}
\alsoaffiliation{Physical and Life Sciences Directorate, Lawrence Livermore National Laboratory, Livermore, California 94550, USA}

\author{Matthew P. Kroonblawd}
\affiliation{Physical and Life Sciences Directorate, Lawrence Livermore National Laboratory, Livermore, California 94550, USA}
\email{kroonblawd1@llnl.gov}

\author{Brenden W. Hamilton}
\affiliation{School of Materials Engineering and Birck Nanotechnology Center, Purdue University, West Lafayette, Indiana, 47907 USA}
\alsoaffiliation{Theoretical Division, Los Alamos National Laboratory, Los Alamos, New Mexico 87545, USA}

\author{H. Keo Springer}
\affiliation{Physical and Life Sciences Directorate, Lawrence Livermore National Laboratory, Livermore, California 94550, USA}

\author{Alejandro Strachan}
\affiliation{School of Materials Engineering and Birck Nanotechnology Center, Purdue University, West Lafayette, Indiana, 47907 USA}
\affiliation{Department of Materials Engineering, Purdue University}
\email{strachan@purdue.edu}

\title{Modeling Framework to Predict Melting Dynamics at Microstructural Defects in TNT-HMX High Explosive Composites}

\abbreviations{HMX,TNT,EM,MD,IFE,SFE}

\begin{document}




\begin{abstract}

Many high explosive (HE) formulations are composite materials whose microstructure is understood to impact functional characteristics. Interfaces are known to mediate the formation of hot spots that control their safety and initiation. To study such processes at molecular scales, we developed all-atom force fields (FFs) for Octol, a prototypical HE formulation comprised of TNT (2,4,6-trinitrotoluene) and HMX (octahydro-1,3,5,7-tetranitro-1,3,5,7-tetrazocine). We extended a FF for TNT and re-casted it in a form that can be readily combined with a well-established FF for HMX. The resulting FF was extensively validated against experimental results and density functional theory calculations. We applied the new combined TNT-HMX FF to predict and rank surface and interface energies, which indicate that there is an energetic driver for coarsening of microstructural grains in TNT-HMX composites. Finally, we assess the impact of several microstructural environments on the dynamic melting of TNT crystal under ultrafast thermal loading. We find that both free surfaces and planar material interfaces are effective nucleation points for TNT melting. However, MD simulations show that TNT crystal is prone to superheating by at least 50~K on sub-nanosecond timescales and that the degree of superheating is inversely correlated with surface and interface energy. The modeling framework presented here will enable future studies on hot spot formation processes in accident scenarios that are governed by strong coupling between microstructural interfaces, material mechanics, momentum and energy transport, phase transitions, and chemistry. 

\end{abstract}


\section{Introduction}

Intentional or accidental initiation of secondary high explosives (HEs) is widely understood to be driven by hot spots, which are regions of localized energy that accelerate chemical reactions.\cite{handley_2018, hamilton_2021} Hot spots are understood to form as a consequence of complicated interactions between a mechanical stimulus, such as a shock wave, and microstructual defects, such as pores or interfaces.\cite{Davis1981, Field1992} Since the earliest computer simulations of hot spot formation in the 1960's,\cite{mader_1965} pore collapse has arguably been the most extensively characterized hot spot mechanism owing to the understanding that it is the dominant mechanism for shock initiation in many secondary HEs (See, for example Refs.~\citenum{Dattelbaum2010, wood_2015, Springer2018, Li2020, Das2021, Duarte2021, Hamilton2021}). Hot spot mechanisms involving dynamically formed interfaces such as shear banding\cite{rimoli_2010, Kroonblawd2020, hamilton_2024} and crack propagation\cite{Dienes2006, Grilli2018} have received attention, but the role of pre-existing material interfaces in hot spot formation is less well understood. Characterizing the role of microstructural surfaces and interfaces on HE initiation is challenging due to small length and time scales associated with localization of energy and rapid releases of energy \cite{Carter2009, Dlott2015}. For example, experimental work found that interfacial delamination at interfaces between a polymer and energetic material is the primary mechanism of hot spot formation under ultrasonic insults.\cite{Dlott2015} While energy localization into hot spots around interfaces and defects is known to contribute to the chemical initiation of HEs, the formation mechanisms are not fully understood.\cite{Dandekar2019} Developing a better physical understanding of the role of interfaces in hot spot formation will help improve computational models to predict HE initiation and would benefit a range of application areas including construction/demolition, mining, defense, and propulsion.\cite{Zhou2016, Ravindran2023}

In many cases, HEs are formulated composite materials comprised of an energetic crystalline phase bound with a matrix phase such as a polymer or a melt-castable HE (i.e., an HE that is chemically stable upon melting).\cite{Gibbs1980,Chen2023} This leads to rich microstructure with many interfaces between the various material phases.\cite{chun_2020} Characterizing surface and interface energies of formulation constituents can help guide efforts to produce HEs with controlled microstructures \cite{Forrest2021,Zhang2009}. Yet, even with such knowledge, better modeling tools are needed to understand how natural or engineered features will impact hot spot formation. Computational modeling in a continuum-mechanics framework at the grain-scale can explicitly resolve the formation of hot spots at realistic microstructural defects,\cite{nguyen_2023} but the physics in these models is necessarily coarse-grained and can be difficult to parameterize and validate. Moreover, continuum models can be prone to model-form errors and uncertainties in the coupling between different physical processes such as crystal mechanics, phase transitions, energy and momentum transport, and chemical reactions.\cite{Das2021, Kroonblawd2021} In contrast, all-atom simulation techniques such as molecular dynamics (MD) can directly model these highly coupled processes with few assumptions, but MD simulations of HEs are generally limited to sub-micron scale domains with current computers.\cite{Kroonblawd2025} Specialized nonreactive MD force field (FF) models have been developed for many pure organic HEs,\cite{Smith1999, sorescu_2000, borodin_2008, Bedrov2009, Neyertz2012, bidault_2019, wang_2019, Kroonblawd2024} which are widely cited and have underpinned numerous MD studies aimed at understanding fundamental properties of HEs. Scale bridging between MD and continuum models has emerged as a means to directly calibrate\cite{wood_2018} and validate\cite{Zhao2020, nguyen_2024, herrin_2024} grain-scale continuum HE models, but these efforts have been largely focused on pure materials and not composites.  Extending MD capabilities to inform multi-material continuum simulations\cite{Das2022, Springer2023} has potential to significantly improve the physical accuracy of models of real HE formulations.

Direct all-atom MD simulations of surfaces and interfaces can help deconvolve separate effects that contribute to hot spot formation and identify new mechanisms that are missing in continuum models. Quasi-1D MD simulations have shown that shocking free surfaces can lead to adiabatic free expansion and recompression and drive hot spots to very high temperatures.\cite{holian_2002} Shocking surfaces with 2D features can induce molecular jetting, which can further enhance recompression work and the peak hot spot temperature as compared to shock interactions with planar surfaces.\cite{Li2020} Shearing forces at planar solid-solid interfaces can also influence hot spot reactivity in pure HEs\cite{islam_2020} and these localization effects arise in multiple different HE materials.\cite{Hamilton2023} Interfaces between different phases of the same material, such as a solid-liquid interface, can induce hot spot formation processes that vary depending on the shock propagation direction.\cite{jiang_2016} Very recent reactive MD simulations have shown that expansion and recompression of a matrix phase can significantly impact (and accelerate) decomposition reactions in an HE.\cite{macatangay_2024} The above studies point to a strong likelihood for additional unusual effects at material interfaces in formulated HEs, especially in systems where the matrix phase is also energetic.

\begin{figure}[t!]
  \centering
  \includegraphics[width=0.98\textwidth]{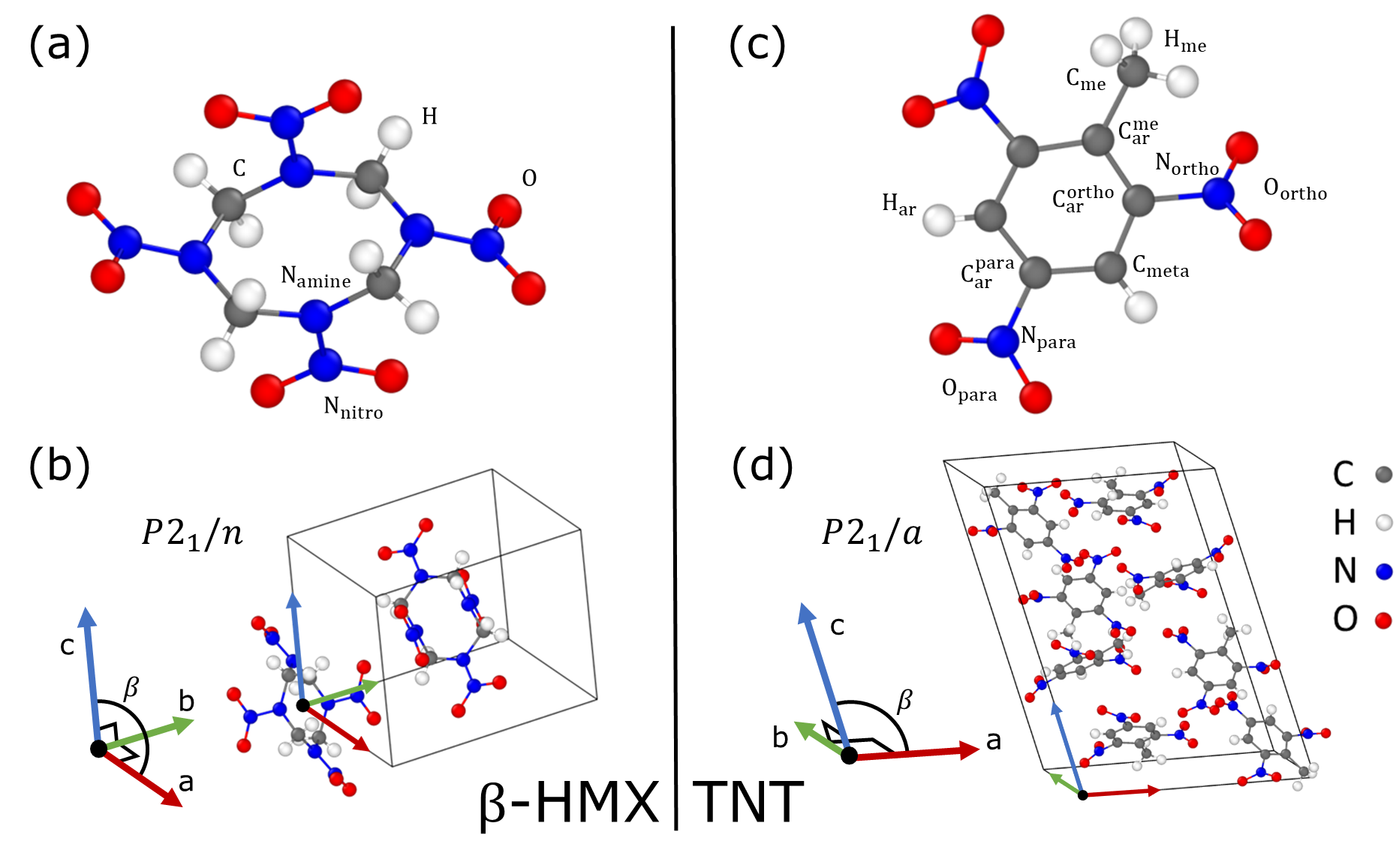}
  \caption{ Snapshots showing the (a) HMX molecule, (b) monoclinic ($P2_1/n$) $\beta$-HMX crystal unitcell, (c) TNT molecule, and (d) the monoclinic ($P2_1/a$) TNT crystal unitcell. Atoms are colored gray, white, blue, and red for C, H, N, and O, respectively. Chemically distinct atom types in the TNT and HMX FFs are labeled with annotations to the molecular structures.}
  \label{fig:unitcells}
\end{figure}

To the above ends, we focus here on developing MD FF models and work flows suitable for performing large-scale simulations of composite HE materials containing TNT (2,4,6-trinitrotoluene) and HMX (octahydro-1,3,5,7-tetranitro-1,3,5,7-tetrazocine). Composites containing TNT and HMX serve as a scientifically interesting and technologically relevant model system that captures the many of the salient features of a multi-material HE formulation. Both TNT and HMX exist as molecular crystalline solids near standard conditions\cite{cady_1963, Vrcelj2003} (see Figure~\ref{fig:unitcells}) and are often combined together in formulations such as Octol.\cite{Lanzerotti1992} Similar formulations such as Composition B replace HMX with its close chemical relative RDX (1,3,5-trinitro-1,3,5-triazinane); HMX and RDX differ chemically only in that RDX is a 6-member ring and HMX is an 8-member ring.\cite{Gibbs1980} In these formulations, TNT acts as an energetic matrix material that facilitates casting HE parts owing to its low melting point and its chemically stability in the liquid phase.\cite{Dattelbaum2014} While equilibrium mixtures involving TNT and nitramines such as HMX and RDX have received recent attention in the context of equation of state modeling,\cite{myint_2016, myint_2017} the dynamic evolution of microstructure in TNT-HMX composites--like other HE formulations--remains poorly understood at an atomistic level. For this reason, we anticipate that developing an MD modeling work flow to simulate TNT-HMX systems would find numerous applications to help elucidate the fundamental physics of hot spot formation at HE-matrix interfaces and in studies focused on specific formulations such as Octol or Composition B.

In what follows, we first extend and calibrate an existing\cite{Neyertz2012} rigid-bond FF for TNT to enable simulations with fully flexible molecules that can be directly combined with the well-validated FF model for HMX and RDX developed by Smith and Bharadwaj.\cite{Smith1999} This new TNT FF is extensively validated against experimental data on the molecular vibration spectrum, the high-pressure equation of state, melting point, coefficient of thermal expansion, and other thermodynamic data. We then validate cross-interactions between the TNT and HMX FFs against density functional theory (DFT) calculations of the potential energy surface for dimer interactions. The combined TNT-HMX FF is then applied in an MD work flow based on the generalized crystal-cutting method\cite{Kroonblawd2016} (GCCM) to predict the energetic stability of crystal surfaces and crystal-crystal interfaces involving TNT and HMX. Finally, we assess the dynamics of TNT melting in plausible microstructural environments including the homogeneous bulk, at free surfaces, and at interfaces with HMX. Our MD simulations show that free surfaces are more effective nucleation points for melting than crystal-crystal interfaces and they provide tentative evidence for a correlation between the surface energy and propensity to superheat.

\section{Methods}
\label{sec:Methods}

\subsection{DFT Calculations}
\label{subsec:methodsDFT}

DFT calculations were performed at the Perdew-Burke-Ernzerhof D3 dispersion correction (PBE-D3) level\cite{perdew_1996, grimme_2010} using the CP2K code\cite{kuhne_2020} to assess the accuracy of the TNT FF and parameterize cross-interactions between the HMX and TNT FFs. The electronic structure was evaluated with Geodecker-Teter-Hutter pseudopotentials\cite{goedecker_1996} and the DZVP-MOLOPT-GTH basis set\cite{vandevondele_2007} using $\Gamma$-point sampling without spin polarization. A 400~Ry cutoff was used for the finest grid and the self-consistent field convergence threshold was set to $1.0 \times 10^{-6}$~{H}.
All calculations were performed with 3D periodic boundary conditions; isolated dimer configurations were prepared using a large $35 \times 20 \times 20$~{\AA}$^{3}$ cell with excess vacuum to reduce interactions across the periodic boundary. Optimizations were performed with a root-mean-square force-based accuracy threshold of $1.0 \times 10^{-3}$~{H/Bohr}.

\subsection{Classical MD Simulation Details}
\label{subsec:methodsMD}

Classical MD simulations of HMX, TNT, and TNT-HMX mixtures were performed using the LAMMPS software package\cite{Thompson2022} (release: 2 Aug. 2023) with the nonreactive FFs described below. Atomic repulsion and dispersion interactions were modeled as pairwise potentials that were evaluated in real space up to an 11~{\AA} cutoff. Electrostatic interactions between atom-centered partial charges were also evaluated in real space using the Wolf potential\cite{wolf_1999} with an 11~{\AA} cutoff and the damping parameter set to 0.2~{\AA}. Partial charges were obtained for the HMX and TNT FFs from Ref \citenum{Smith1999} and Ref \citenum{Neyertz2012}, respectively. All intramolecular 1-2 and 1-3 nonbonded interactions were excluded and no scaling was applied to 1-4 interactions. LAMMPS input files for example HMX, TNT, and TNT-HMX composite systems are provided in the Supporting Information.

Isobaric-isothermal (NPT) and isochoric-isothermal (NVT) used a Nos\'e-Hoover-style thermostat and barostat with respective damping parameters of 0.1 ps and 1.0 ps. Isobaric simulations involving crystalline phases were performed using a triclinic barostat and those involving only liquid phases were performed with an isotropic barostat. All trajectories were integrated with a 0.5~fs timestep.

Calculations of bulk material properties were performed using standard MD methods that are described at relevant points in the text. Properties of pure crystalline phases such as the equation of state were calculated using configurations generated by replicating a crystal unitcell. Analogous calculations on pure liquid phases were performed using a cubic simulation cell obtained by melting a crystal supercell that contained 400 molecules, which is sufficiently large to avoid finite size effects in computed structural and transport properties of molecular organic HEs \cite{Kroonblawd2023}. Some simulations involving more complicated configurations, such as those with exposed crystal facets or multi-material interfaces, were prepared using the GCCM code.\cite{Kroonblawd2016} All snapshots of system configurations were prepared using the OVITO graphical user interface and visualization software.\cite{ovito}

\subsection{HMX Force Field}
\label{subsec:methodsHMXFF}

The HMX FF used in this work was based on the widely adopted model by Smith and Bharadwaj developed in 1999.\cite{Smith1999} This FF is split into two contributions to the total potential energy, namely the bonded $U_{\rm{b}}$ and nonbonded $U_{\rm{nb}}$ interactions as
\begin{equation}
U_{\rm{tot}} = U_{\rm{b}} + U_{\rm{nb}}.
\label{eq:tot_nrg}    
\end{equation}The bonded interactions were modeled as a sum over potential energy functions describing 2-center bond stretching, 3-center angle bending, and 4-center proper and improper dihedral torsion within the molecule as
\begin{equation}
    U_{\rm{b}} = \sum_{\rm{bonds}}U_{\rm{bnd}}(r_{ij}) + \sum_{\rm{angles}} U_{\rm{ang}}(\theta_{ijk}) + \sum_{\rm{dihedrals}} U_{\rm{dih}}(\varphi_{ijkl}) + \sum_{\rm{impropers}} U_{\rm{imp}}(\chi_{ijkl}).
\label{eq:HMX_bonded_nrg}
\end{equation}Bond terms were modeled as harmonic functions using the equilibrium lengths from Ref.~\citenum{Smith1999} and with the force constants described in Ref.~\citenum{Kroonblawd2016}. Angle and improper dihedral terms were also modeled as harmonic functions using the original parameters.\cite{Smith1999} Proper dihedrals were modeled as cosine series. The full set of bonded interaction functions and parameters are given in the Supporting Information.

\begin{table}[t!]
\centering
\caption{HMX FF Nonbonded Parameters}
\begin{tabular}{ccccc}
\hline \hline 
Atom Type & $q$ (e) & $A$ (kcal mol$^{-1}$) & $\rho$ ({\AA}$^{-1}$) & $C$ (kcal mol$^{-1}$ {\AA}$^6$) \\
\hline 
H & 0.270000  & 2649.7 & 0.26738  & 27.40 \\
\(\mathrm{N_{nitro}}\) & 0.860625 & 60833.9 & 0.26455 & 500.0 \\
\(\mathrm{N_{amine}}\) & 0.056375 & 60833.9 & 0.26455 & 500.0 \\
O & -0.458500 & 75844.8 & 0.24612 & 398.9 \\
C & -0.540000 & 14976.0 & 0.32362 & 640.8 \\
\hline \hline 
\end{tabular}
\label{tab:HMX-nb}
\end{table}

Nonbonded pairwise interactions in the HMX FF were modeled using the Buckingham potential (exp-6) with Coulombic interactions and an extra $r^{-12}$ term to compensate for the instability in the Buckingham potential at short separation distances. This function, 
\begin{equation}
U_{\text{nb}}^{\text{exp-6-12}} (r) = \sum_{\text{pairs}}  A_{\alpha \beta} \cdot \exp \left( \frac{-r}{\rho_{\alpha \beta}} \right) - C_{\alpha \beta} \cdot r^{-6}  + D \left( \frac{12 \cdot \rho_{\alpha \beta}}{r} \right)^{12} + \frac{q_{\alpha} q_{\beta}}{4 \pi \epsilon_0 r},
\label{eq:nonbonded}
\end{equation}differs from the original 1999 form, but was originally proposed for TATB\cite{Bedrov2009} and was subsequently extended to HMX\cite{Pereverzev2020}. Parameters are listed in Table~\ref{tab:HMX-nb} and include the modified partial charges developed in 2001 by Bedrov et al.\cite{Bedrov2001} Similar to TATB, we set \(D = 5 \times 10^{-5}\) kcal/mol for all interaction pairs. Heteroatom interaction parameters $A_{\alpha\beta}$, $\rho_{\alpha\beta}$, and $C_{\alpha\beta}$ were determined from the homoatom parameters $A_{\alpha\alpha}$, $\rho_{\alpha\alpha}$, and $C_{\alpha\alpha}$ as geometric means.  The exp-6-12 functional form in Equation~\ref{eq:nonbonded} was implemented in LAMMPS using the \textit{hybrid overlay} feature.

\subsection{TNT Force Field}
\label{subsec:methodsTNTFF}

We developed a fully-flexible TNT FF that was based on a rigid-bond FF for TNT originally proposed by Neyertz et al.\cite{Neyertz2012} Bonded interactions in the original TNT FF used cosine-squared functions for angle bending, cosine series for proper dihederals, and a distance-based harmonic function for improper out-of-plane deformations. Nonbonded interactions in the original TNT FF were modeled as pairwise Lennard-Jones 12-6 potentials with Coulomb interactions between fixed atom-centered partial charges. In this report, we extend and re-parameterize the TNT FF with the aims of adding bond stretching for fully flexible molecules, ensuring compatibility with LAMMPS, and bringing its functional form into close compatibility with the HMX FF.

\begin{table}[t!]
\resizebox{\textwidth}{!}{
    \centering
    \caption{TNT FF Bonded Parameters}
    \begin{tabular}{ccccc}
        \hline \hline 
        \multicolumn{5}{c}{Bonds} \\
		Type & \multicolumn{2}{c}{$k_{\rm{bnd}}$ (kcal mol$^{-1}$ \AA$^{-2}$)} & \multicolumn{2}{c}{$r_0$ (\AA)} \\
		\hline 
        \(\mathrm{C_{ar}}\)--\(\mathrm{C_{met}}\) & \multicolumn{2}{c}{500.00} & \multicolumn{2}{c}{1.506} \\
        \(\mathrm{C_{ar}}\)--\(\mathrm{C_{ar}}\) & \multicolumn{2}{c}{500.00} & \multicolumn{2}{c}{1.393} \\
        \(\mathrm{C_{ar}}\)--N & \multicolumn{2}{c}{300.00} & \multicolumn{2}{c}{1.483} \\
        N--O & \multicolumn{2}{c}{880.00} & \multicolumn{2}{c}{1.222} \\
        \(\mathrm{C_{met}}\)--H & \multicolumn{2}{c}{740.00} & \multicolumn{2}{c}{1.090} \\
        \(\mathrm{C_{ar}}\)--H & \multicolumn{2}{c}{800.00} & \multicolumn{2}{c}{1.081} \\
        \midrule
        \multicolumn{5}{c}{Angles} \\
        Type & \multicolumn{2}{c}{$k_{\rm{ang}}$ (kcal mol$^{-1}$ rad$^{-2}$)} & \multicolumn{2}{c}{$\theta_0$ (deg.)} \\
        \hline 
        \(\mathrm{C_{met}}\)--\(\mathrm{C_{ar}}\)--\(\mathrm{C_{ar}}\) & \multicolumn{2}{c}{180.00} & \multicolumn{2}{c}{122.47} \\
        \(\mathrm{C_{ar}}\)--\(\mathrm{C_{ar}}\)--\(\mathrm{C_{ar}}\) & \multicolumn{2}{c}{190.00} & \multicolumn{2}{c}{120.00} \\
        \(\mathrm{C_{ar}}\)--\(\mathrm{C_{ar}}\)--N & \multicolumn{2}{c}{190.00} & \multicolumn{2}{c}{118.60} \\
        \(\mathrm{C_{ar}}\)--\(\mathrm{C_{ar}}\)--H & \multicolumn{2}{c}{60.00} & \multicolumn{2}{c}{120.54} \\
        \(\mathrm{C_{ar}}\)--N--O & \multicolumn{2}{c}{200.00} & \multicolumn{2}{c}{117.30} \\
        O--N--O & \multicolumn{2}{c}{230.00} & \multicolumn{2}{c}{125.39} \\
        H--\(\mathrm{C_{met}}\)--H & \multicolumn{2}{c}{50.00} & \multicolumn{2}{c}{108.06} \\
        \(\mathrm{C_{ar}}\)--\(\mathrm{C_{met}}\)--H & \multicolumn{2}{c}{100.00} & \multicolumn{2}{c}{110.85} \\
        \midrule
        \multicolumn{5}{c}{Proper Dihedrals} \\
        Type & {$k_{{\rm{dih}},0}$ (kcal mol$^{-1}$)} & {$k_{{\rm{dih}},1}$ (kcal mol$^{-1}$)} & {$k_{{\rm{dih}},2}$ (kcal mol$^{-1}$)} & {$k_{{\rm{dih}},3}$ (kcal mol$^{-1}$)} \\
        \hline
        \(\mathrm{H_{met}}\)--\(\mathrm{C_{met}}\)--\(\mathrm{C_{ar}}\)--\(\mathrm{C_{ar}}\) & 0.1195 & -0.3609 & 0.0 &  0.4804 \\
        \(\mathrm{C_{ar}}\)--\(\mathrm{C_{ar}}\)--\(\mathrm{N_{ortho}}\)--O & 1.979 & 0.0 & -1.979 & 0.0 \\
        \(\mathrm{C_{ar}}\)--\(\mathrm{C_{ar}}\)--\(\mathrm{N_{para}}\)--\(\mathrm{O}\) & 3.200 & 0.0 & -3.200 & 0.0 \\
        *--\(\mathrm{C_{ar}}\)--\(\mathrm{C_{ar}}\)--* & 4.000 & 0.0 & -4.000 & 0.0 \\
        \hline 
        \multicolumn{5}{c}{Improper Dihedrals} \\
        Type & \multicolumn{4}{c}{$k_{\rm{imp}}$ (kcal mol$^{-1}$)} \\
        \hline 
        C--C--*--\(\mathrm{C_{ar}^*}\) & \multicolumn{4}{c}{36.5} \\
        \(\mathrm{C_{ar}}\)--O--O--\(\mathrm{N^*}\) & \multicolumn{4}{c}{89.3} \\
        \hline \hline 
    \end{tabular}
\label{tab:TNT-cov}
}
\end{table}

As with the HMX FF, our new TNT FF adopts harmonic functions for bonds, angles, and improper dihedrals and uses cosine power series for proper dihedrals, 
\begin{equation}
  U_{\rm{bnd}}(r) = \frac{1}{2} k_{\rm{bnd}} \left( r - r_0 \right)^2,
  \label{eq:harmonicBnd}
\end{equation} 
\begin{equation}
  U_{\rm{ang}}(\theta) = \frac{1}{2} k_{\rm{ang}} \left( \theta - \theta_0 \right)^2,
  \label{eq:harmonicAng}
\end{equation} 
\begin{equation}
   U_{\rm{dih}}(\phi) = \sum_{n} k_{{\rm{dih}},n} \cos^{n} \varphi,
  \label{eq:cosineDih}
\end{equation} 
\begin{equation}
  U_{\rm{imp}} = \frac{1}{2} k_{\rm{imp}} \chi^2.
  \label{eq:harmonicImp}
\end{equation}Parameters for these functions are listed in Table~\ref{tab:TNT-cov}. As will be discussed extensively in Section~\ref{subsec:TNT-cov}, only the proper dihedral parameters were directly taken from the original TNT FF. Improper dihedral terms were taken from the Bedrov FF for the related nitroaromatic molecule TATB\cite{Bedrov2009} (1,3,5-triamino-2,4,6-trinitrobenzene) and the bond and angle terms were hand-tuned to approximately reproduce TNT vibrational normal mode frequencies and mode characters.

\begin{table}[t!]
\centering
\caption{TNT FF Nonbonded Parameters}
\begin{tabular}{cccccc}
\hline \hline 
Atom Type & $q$ (e) & $A$ (kcal mol$^{-1}$) & $\rho$ ({\AA}$^{-1}$) & $C$ (kcal mol$^{-1}$ {\AA}$^6$) \\
\hline 
\(\mathrm{O_{ortho}}\) & -0.3978 & 102813.2750 & 0.240434435 & 324.9027922 \\
\(\mathrm{O_{para}}\) & -0.3736 & 102813.2750 & 0.240434435 & 324.9027922 \\
\(\mathrm{N_{para}}\) & 0.622 & 70882.76830 & 0.225111834 & 150.8880585 \\
\(\mathrm{C_{ar}^{met}}\) & 0.1657 & 79013.22139 & 0.246873188 & 292.5954464 \\
\(\mathrm{C_{ar}^{ortho}}\) & -0.0548 & 79013.22139 & 0.246873188 & 292.5954464 \\
\(\mathrm{C_{ar}^{meta}}\) & -0.1267 & 79013.22139 & 0.246873188 & 292.5954464 \\
\(\mathrm{C_{ar}^{para}}\) & 0.0826 & 79013.22139 & 0.246873188 & 292.5954464 \\
\(\mathrm{H_{ar}}\) & 0.1693 & 30969.63495 & 0.217858049 & 54.16291046 \\
\(\mathrm{C_{met}}\) & -0.4095 & 79013.22139 & 0.246873188 & 292.5954464 \\
\(\mathrm{N_{ortho}}\) & 0.7218 & 70217.54941 & 0.225111834 & 149.4720079 \\
\(\mathrm{H_{met}}\) & 0.1528 & 58761.00187 & 0.217858049 & 102.7673361 \\
\hline \hline 
\end{tabular}
\label{tab:TNT-nb}
\end{table}

Nonbonded interactions in the TNT FF were recast from their original Lennard-Jones (12-6) form to the same modified Buckingham (exp-6-12) form used by the HMX FF (Equation~\ref{eq:nonbonded}). As will be discussed in Section~\ref{subsec:TNT-nb}, there are multiple standard transformations between the Lennard-Jones and Buckingham potential forms and we chose one that best reproduced the experimental pressure-volume response of TNT crystal. A complete set of partial charges and homoatom parameters $A_{\alpha\alpha}$, $\rho_{\alpha\alpha}$, and $C_{\alpha\alpha}$ are given in Table~\ref{tab:TNT-nb}. As with the HMX FF, heteroatom interaction pairs were computed using geometric means. The $r^{-12}$ parameter $D$ was set to $5 \times 10^{-5}$~kcal/mol for all TNT interaction pairs.

\subsection{Composite TNT-HMX Force Field}
\label{subsec:methodsCombineFF}

Mixtures of HMX and TNT were modeled using a FF that combined the FFs for pure HMX and TNT described just above. Because both pure-material FFs adopt the same nonbonded potential form, cross-terms are straightforward to define in terms of mixing rules. As will be described in Section~\ref{sec:cross-terms}, we used DFT calculations of dimer interaction energies to explore several choices for the mixing rules and identified geometric means as an accurate way to compute heteroatom interaction terms $A_{\alpha\beta}$, $\rho_{\alpha\beta}$, and $C_{\alpha\beta}$ from the homoatom terms. The parameter $D$ in Equation~\ref{eq:nonbonded} was set to $5 \times 10^{-5}$~kcal/mol for all TNT-HMX interaction pairs.

\section{TNT FF Development and Validation} 
\label{sec:SectionFF}

\subsection{Vibrational Spectrum} 
\label{subsec:TNT-cov}

Molecular vibrations influence many properties and processes that govern HE safety and performance, including transport properties such as thermal conductivity \cite{Kumar2019, Kroonblawd2013, Perriot2021, Li2023}, intramolecular vibrational relaxation \cite{Zhang2022, Cole-Filipiak2020_L, Selezenev2008}, and the closely related concept of phonon up-pumping \cite{Hong1996, Hang2017, michalchuk_2021, Bidault2022}. Ensuring these processes are accurately captured in MD simulations involving TNT was one of our main motivations in extending the Neyertz et al. rigid-bond TNT FF to model fully flexible molecules. Vibrational accuracy is strongly coupled to the parameterized potential energy functions for bonded interactions. For this reason, we systematically calibrated the bond and angle function parameters (Eqs.~\ref{eq:harmonicBnd} and \ref{eq:harmonicAng}) so that our fully flexible TNT FF accurately reproduced vibrational normal mode frequencies and displacements of an isolated molecule as predicted by DFT.

An initial set of potential energy function parameters was specified that drew from multiple sources of information, including the rigid-bond Neyertz et al. parameterization and from FFs for other organic HEs, including a variant\cite{Kroonblawd2016} the HMX FF due to Smith and Bharadwaj \cite{Smith1999} and the fully flexible version\cite{Kroonblawd2013} of the TATB FF by Bedrov et al.\cite{Bedrov2009} The rigid-bond TNT force field constrains bond distances to equilibrium values, which we used as the $r_{0}$ values in Equation \ref{eq:harmonicBnd}. Starting points for the bond force constants $k_{\rm bnd}$ were taken from the HMX and TATB FFs. Angle bending and proper dihedral terms were directly ported from the rigid-bond TNT FF and improper dihedral terms were taken from the TATB FF. The proper and improper dihedral terms were left unmodified in our calibration as these terms were originally fit to electronic structure theory calculations and largely determine low-frequency vibrational modes that involve many coupled internal coordinates.

\begin{figure*}[t!]
  \centering
  \includegraphics[width=6.5in]{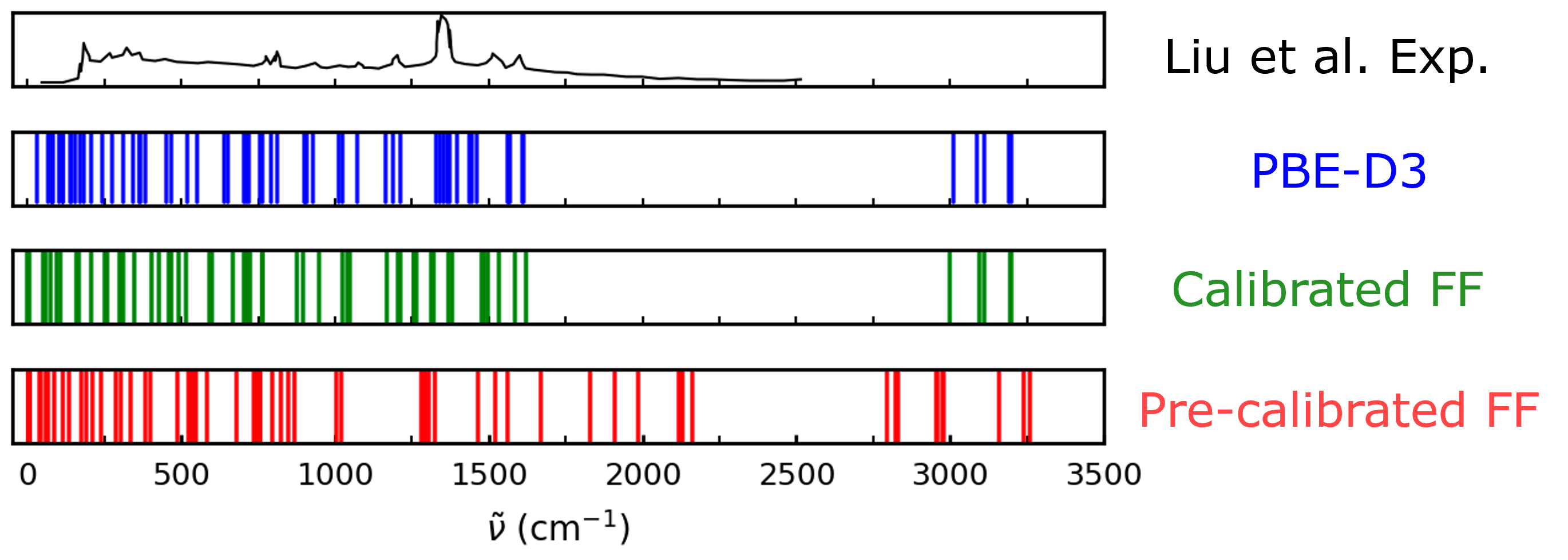}
  \caption{Vibrational spectra for TNT, including the Raman experiments by Liu et al.\cite{Liu2017} and normal mode frequencies for an isolated TNT molecule predicted by a PBE-D3 DFT calculation and with the pre-calibrated and calibrated TNT FF.}
  \label{fig:VNM-spectra}
\end{figure*}

Figure \ref{fig:VNM-spectra} shows the normal mode frequency spectrum predicted by our reference PBE-D3 DFT calculation and the pre-calibrated TNT FF in comparison to results from a Raman experiment\cite{Liu2017} and from our final calibrated FF. As can be clearly seen, the pre-calibrated FF exhibits many modes with frequencies between 1700~cm$^{-1}$ and 3000~cm$^{-1}$ that are inconsistent with DFT and experiment. Similar to previous work on the TATB FF,\cite{Kroonblawd2013} we calibrated a set of bond and angle force constants by hand through an iterative process starting with the highest-frequency motions. This calibration used the normal mode displacements as a guide to identify internal coordinates associated with particular frequencies.

The five highest frequency modes near and above 3000~cm$^{-1}$ correspond to C-H stretching motions that were straightforward to fit because they are largely decoupled from the other vibrational modes. These include nearly degenerate symmetric and asymmetric aromatic hydrogen stretching modes near 3200~cm$^{-1}$ and the symmetric and asymmetric methyl hydrogen stretching modes respectively at 2999~cm$^{-1}$ and near 3100~cm$^{-1}$. Lower frequency modes ($<$1700~cm$^{-1}$) were more complicated as they involved multiple internal coordinates. Fitting these modes required simultaneous adjustments to the C-C, C-N, and N-O bond force constants and modifications to all of the angle force constants. We note that our flexible FF has significantly softer angle bending force constants than the rigid-bond FF by Neyertz et al., which is similar to previous extensions of a rigid-bond FF for TATB.\cite{Kroonblawd2013} We ultimately attempted to fit modes with frequencies above 1000~cm$^{-1}$.

While the character of individual modes was examined during calibration, the overall accuracy of all the modes was assessed using a mode correlation matrix computed as
\begin{equation}
  X_{ij} = | \mathbf{N}_{i}^{\rm FF} \cdot \mathbf{N}_{j}^{\rm DFT} |^2.
  \label{eq:modeprojection}
\end{equation}Here, $\mathbf{N}_{i}^{\rm FF}$ and $\mathbf{N}_{j}^{\rm DFT}$ are length-63 unit eigenvectors that encode the mode displacements for the $i^{\rm th}$ FF mode and $j^{\rm th}$ DFT mode in mass-weighted Cartesian coordinates expressed in equivalent rotational frames. Thus defined, a value of unity indicates perfect unsigned correlation between a given pair of FF and DFT modes and a value of zero indicates no correlation.

\begin{figure*}[t!]
  \centering
  \includegraphics[width=6.5in]{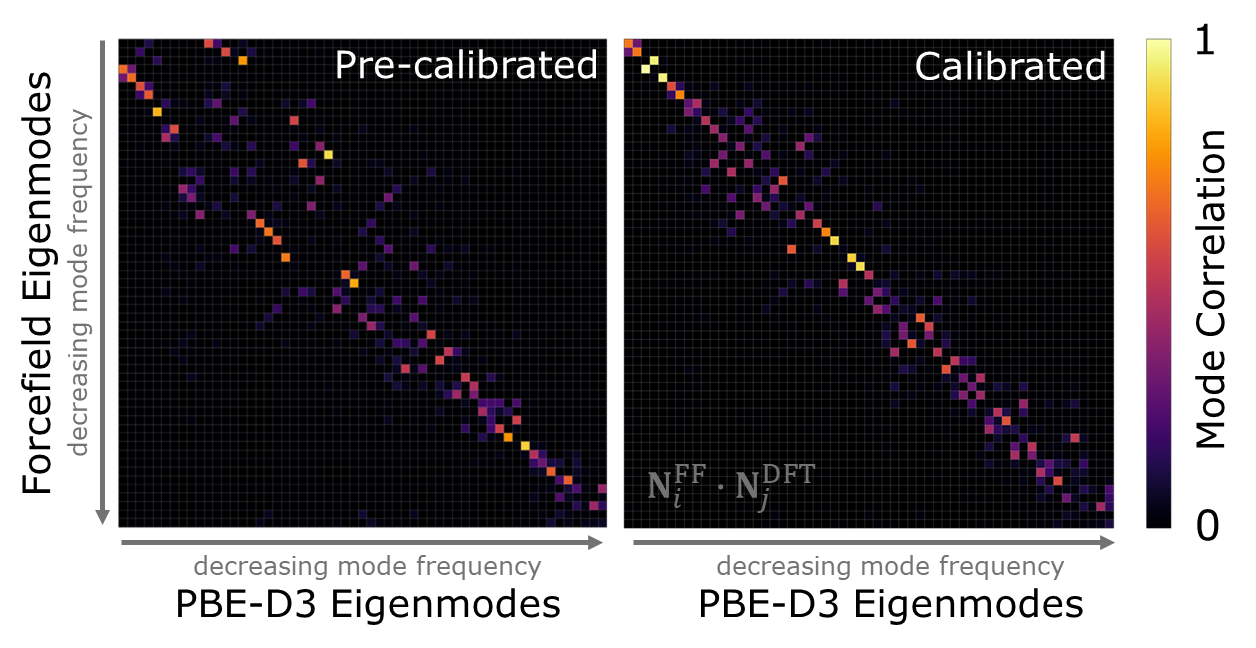}
  \caption{Vibrational normal mode correlation matrices computed using Equation~\ref{eq:modeprojection} for the (left) pre-calibrated TNT FF and (right) calibrated TNT FF.}
  \label{fig:VNM-character}
\end{figure*}

Resulting mode correlation matrices for the pre-calibrated and calibrated FFs are given in Figure~\ref{fig:VNM-character}. These plots compare the $3N - 6 = 57$ vibrational normal modes of an isolated TNT molecule. Perfect agreement between the FF and DFT in terms of both mode character and frequency ordering of the modes would result in a perfect diagonal, that is $X_{ij} = 1$ if $i = j$ and $X_{ij} = 0$ otherwise. The off-diagonal scatter in both plots indicates that there are some differences between the FF- and DFT-predicted modes. Tighter clustering along the diagonal for the calibrated FF case shows that our mode-by-mode calibrations result in a significant improvement in the vibrational accuracy of the TNT FF. This is particularly evident for the highest-frequency modes in the upper left-hand corner of the matrix plot. The lowest frequency modes in the lower right-hand corner of the matrix are well described by the pre-calibrated set of force constants, which is likely because the proper and improper dihedral force constants were already tuned to reproduce electronic structure theory calculations. Our calibrated FF largely retains good accuracy of the mode character and frequency ordering for these low-frequency motions while significantly improving the high frequency region.

\subsection{PV Isotherm Near Room Temperature} 
\label{subsec:TNT-nb}

The original rigid-bond TNT FF by Neyertz et al.\cite{Neyertz2012} uses a Lennard-Jones plus Coulomb interaction potential 
\begin{equation}
    U_{\text{nb}}^{\text{LJ}} (r) = \sum_{pairs}^{}{4\varepsilon_{\alpha\beta}\left\lbrack \left( \frac{\sigma_{\alpha\beta}}{r_{}} \right)^{12} - \left( \frac{\sigma_{\alpha\beta}}{r_{}} \right)^{6} \right\rbrack} + \frac{q_{\alpha} q_{\beta}}{4 \pi \epsilon_0 r},
\label{eq:nb_lj}
\end{equation}to describe nonbonded interactions between atoms. Parameters $\varepsilon_{\alpha\beta}$ and $\sigma_{\alpha\beta}$ were empirically adjusted by Neyertz et al. to ``maintain the crystalline structures'' of the monoclinic and orthorhombic phases and the partial charges were tuned to reproduce experimental densities of solid and liquid phases along with the sublimation enthalpy. We recast the TNT nonbonded interactions in terms of the exp-6 plus Coulomb potential 
\begin{equation}
    U_{\text{nb}}^{\text{exp-6}} (r) = \sum_{\text{pairs}}  A_{\alpha \beta} \cdot \exp \left( \frac{-r}{\rho_{\alpha \beta}} \right) - C_{\alpha \beta} \cdot r^{-6} + \frac{q_{\alpha} q_{\beta}}{4 \pi \epsilon_0 r},
\label{eq:nb_exp-6}
\end{equation}to facilitate the derivation of mixing rules to describe TNT-HMX cross interactions. This conversion was performed using the approach described in Ref.~\citenum{Mayo1990}, which introduces a parameter $\zeta$ to derive a set of equations 
\begin{equation}
    A = \varepsilon\left( \frac{6}{\zeta - 6} \right)e^{\zeta},
\label{eq:A}
\end{equation}
and
\begin{equation}
    \rho = \frac{1}{B} = \frac{2^{\frac{1}{6}}\sigma}{\zeta},
\label{eq:rho}
\end{equation}
and
\begin{equation}
    C = \left( \frac{\zeta}{\zeta - 6} \right)2\sigma^{6}\varepsilon,
\label{eq:C}
\end{equation}that relate $\varepsilon$ and $\sigma$ to $A$, $\rho$, and $C$. Two common choices for $\zeta$ are $\zeta = 12.0$, which matches long-range attraction between the two forms, and $\zeta = 13.772$, which matches curvature around the equilibrium bond distance.\cite{Mayo1990} A third option with $\zeta = 12.653$ matches the integral of the energy from the equilibrium bond distance to infinite separation.\cite{Lim2009} We chose from the three options for $\zeta$ based on a comparison against experimental diamond anvil cell (DAC) data from Stevens et al.\cite{Stevens2008} and Bowden et al.\cite{Bowden2014} for the isothermal pressure-volume response of the stable monoclinic TNT crystal polymorph at room temperature. Both Stevens et al. and Bowden et al. performed quasihydrostatic experiments using a pressure transmitting medium and nonhydrostatic experiments performed without a transmitting medium. Our validation testing focuses primarily on the quasihydrostastic data.

\begin{figure}[t!]
  \centering
  \includegraphics[width=3.35in]{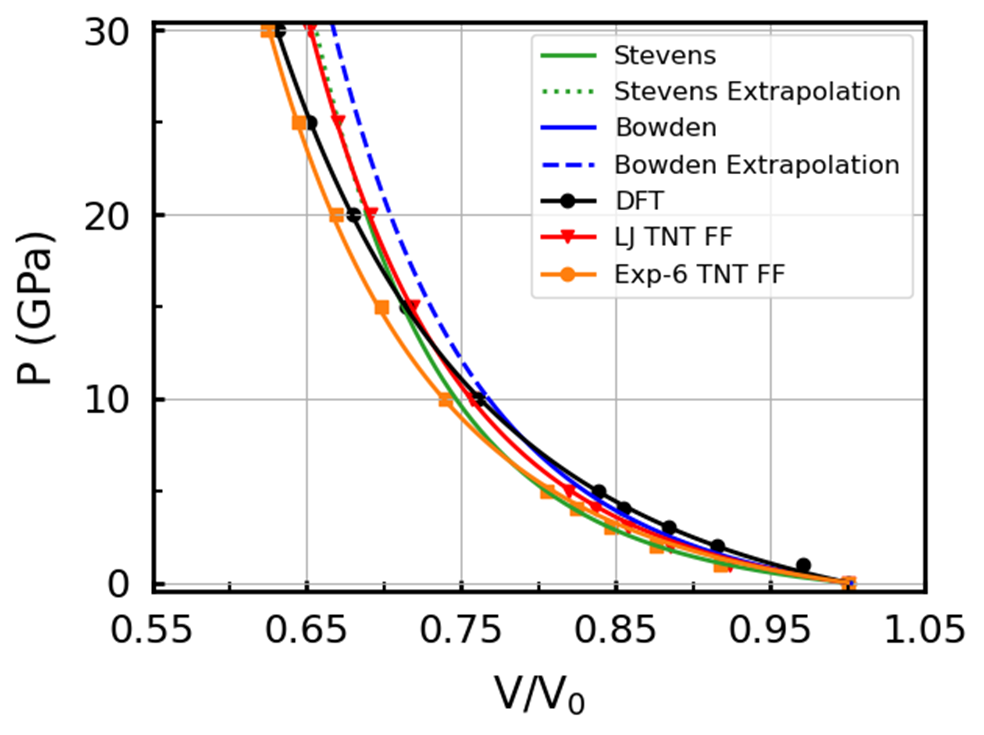}
  \caption{Comparison of isothermal $P(V)$ curves for monoclinic TNT crystal polymorph obtained from the fully flexible TNT FF using the Lennard-Jones potential form, the TNT FF with the exp-6 form using $\zeta = 13.772$, PBE-D3 DFT calculations, and the hydrostatic experimental results from Stevens et al.\cite{Stevens2008} and Bowden et al.\cite{Bowden2014} Symbols correspond to calculated values and curves to Murnaghan fits. Extrapolations of Murnaghan fits beyond the fitted pressure interval are plotted as dashed curves.}
  \label{fig:PV-EOS}
\end{figure}

Figure~\ref{fig:PV-EOS}(a) shows the predicted $P(V)$ curve obtained with the fully flexible TNT FF at 300~K using the original LJ potential, the TNT FF using the exp-6 potential with our final choice for $\zeta$ (13.772), the 0~K isotherm predicted by PBE-D3 DFT calculations, and the room temperature quasihydrostatic experiments by Stevens et al. and Bowden et al. Plots comparing the other choices for $\zeta$ and with the nonhydrostatic experimental data are provided in the SI and show that smaller $\zeta$ leads to $P(V)$ responses that are much too compressible, even at low pressure below 5~GPa. Results for the TNT FF were obtained with a $3 \mathbf{a} \times 6 \mathbf{b} \times 2 \mathbf{c}$ supercell and the DFT results were obtained with a $1 \mathbf{a} \times 2 \mathbf{b} \times 1 \mathbf{c}$ supercell. All calculations were performed at selected pressure states of 0, 1, 3, 5, 10, 15, 20, 25, and 30~GPa.

Focusing first on the experimental data, it is clear that there are differences in the two $P(V)$ curves, even below 10~GPa where both studies obtained direct measurements. The Stevens et al. experiment went up to 26.5~GPa and they observed a solid-solid phase transition to the orthorhombic phase at 20~GPa. Our DFT predictions are quite similar to the data from Bowden et al. up to their highest measured pressure (10~GPa), but are much more compressible above 10~GPa than would be expected based on extrapolations of the experimental data. Visual inspection shows that the LJ version of the TNT FF yields a $P(V)$ response that falls between the bounds of both experiments up to 20~GPa. The exp-6 version is very similar to the Stevens et al. experiment up to $\approx$7~GPa and is modestly more compressible above that pressure. Overall, the exp-6 TNT FF version is more compressible than the LJ version over the entire pressure interval, but both versions are consistent with experiment below 7~GPa. Based on these results, MD studies at higher pressures would likely be more accurate using the LJ version as compared to the exp-6 version.

\begin{table}[t!]
    \centering
    \caption{Murnaghan fit parameters for the monoclinic TNT polymorph}
    \begin{tabular}{lcccr}
    \hline \hline 
     & $V_0$ & $K_0$ & $K'_0$ & Refs. 	\\
     & ({\AA}$^3$) & (GPa) & (Unitless) & \\     
    \hline
    Hydrostatic results & & & &  \\
    Stevens et al.  					&  
    1828.8		 		         		&     
    \( 8.52 \)      					& 
    \( 8.0 \)              				&
    \citenum{Stevens2008}				\\ 
    Bowden et al. 		 				&  
    1828.8	 		         			&     
    \(12.8 \pm 1.4\)    							& 
    \(7.1 \pm 0.8 \)           						&
    \citenum{Bowden2014}				\\            
    Flexible TNT FF (LJ Form)	     	& 
    1868.3 			    				&       
    12.1			        				&
    6.74           						&
    This work							\\  
    Flexible TNT FF (exp-6 Form)		& 
    1832.8 			    				&       
    11.8			        				&
    5.92               					&
    This work                  			\\
    PBE-D3              	            &
    1822.2	 			    				&       
    18.0			        				&
    4.75              					&
    This work                  			\\      
    \hline    
    Nonhydrostatic results & & & &  \\   
    Stevens et al.  					&  
    1828.8		 		         		&     
    \(7.3 \pm 0.5\)     							& 
    \(9.6 \pm 0.4\)             					&
    \citenum{Stevens2008, Bowden2014}	\\                 
    Bowden et al. (Exp. 1)				&
    1828.8		 		         		&     
    \( 10.8 \pm 0.6 \)      			& 
    \( 6.6 \pm 0.4\)             		&
    \citenum{Bowden2014} 				\\    
    Bowden et al. (Exp. 2)				&
    1828.8		 		         		&     
    \( 9.8 \pm 0.4\)      				& 
    \( 8.2 \pm 0.3\)             		&
    \citenum{Bowden2014} 				\\
    Bowden et al. (Exp. 3)      		&
    1828.8		 		         		&     
    \( 9.6 \pm 0.9 \)      				& 
    \( 8.0 \pm 0.6 \)              		&
    \citenum{Bowden2014} 				\\
    \hline \hline 
    \end{tabular}
    \label{tab:TNT-bulk-modulus}
\end{table}

To better quantify the accuracy of the TNT FF $P(V)$ response, we fit the predicted $P(V)$ data to the isothermal Murnaghan equation of state 
\begin{equation}
  P^{\rm Murn}(V) = \frac{K_0}{K'_0} \left[ \left( \frac{V}{V_0} \right)^{-K'_0} - 1 \right],
  \label{eq:murnaghan}
\end{equation}which is expressed in terms of $K_0$, $K'_0$, and $V_0$, which are respectively the $P = 1$~atm values for the bulk modulus, the pressure derivative of the bulk modulus, and the unit cell volume. Values for $K_0$, $K'_0$, and $V_0$ obtained from experiments, both forms of the TNT FF, and the DFT calculations are listed in Table~\ref{tab:TNT-bulk-modulus}. We obtained $K_0$ and $K'_0$ for the TNT FFs and DFT through a nonlinear least-squares fit in which we fixed $V_0$ to the value predicted by each model at ambient pressure (0 GPa). Uncertainties were calculated from the square root of the associated diagonal element in the resulting covariance matrix.

\begin{figure}[t!]
  \centering
  \includegraphics[width=3.35in]{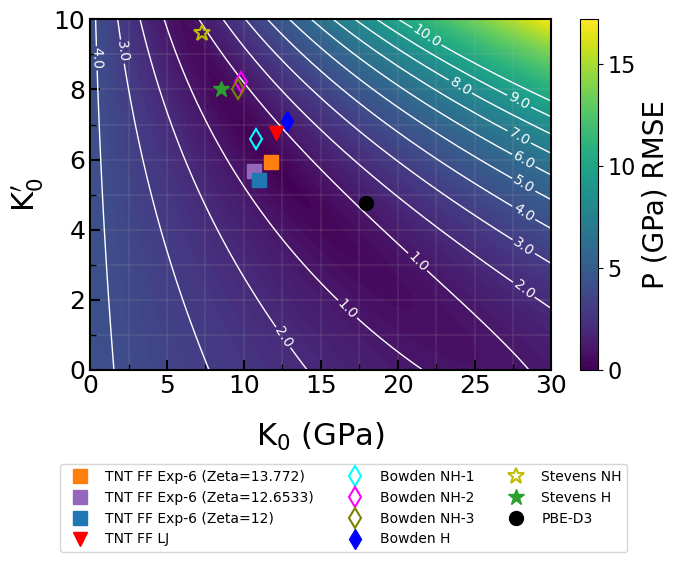}
  \caption{Contour map showing the RMSE of the predicted $P(V)$ response of the exp-6 TNT FF at 300~K plotted as a function of Murnaghan fit parameters $K_0$ and $K_{0}'$ with fixed $V_0 = 1832.8$~{\AA}$^3$. Best-fit $K_0$ and $K_{0}'$ values for the experiments of Stevens et al.\cite{Stevens2008} and Bowden et al.\cite{Bowden2014}, for the LJ form of the TNT FF, and the DFT calculations are also shown for comparison. Filled symbols correspond to hydrostatic results and open symbols to nonhydrostatic results.}
  \label{fig:PV-EOS-error}
\end{figure}

As can be seen in the table, both versions of the TNT FF exhibit a bulk modulus at ambient pressure that is consistent the Bowden et al. hydrostatic experiment within propagated uncertainty. The pressure derivative $K'_0$ of the exp-6 FF version is smaller and outside experimental uncertainty, which tracks with its greater compressibility at high pressure as seen in the $P(V)$ curves. Among the three models, DFT is most unlike either hydrostatic experiment, having the largest $K_0$ and smallest $K_{0}'$ among the three models. There is considerable spread in the experimental data, especially considering the nonhydrostatic experiments. Recent characterizations of the pressure distribution within a DAC sample chamber without a transmitting medium show variations by up to 50\% for target pressures below 10~GPa,\cite{jiang_2022} so it is plausible that the various experiments measured different non-hydrostatic environments.

While useful for comparison, focusing on the fit parameters that correspond to ambient pressure provides only limited insight into the accuracy of the various models. Figure~\ref{fig:PV-EOS-error} shows the root mean-square error (RMSE) surface for fitting the exp-6 TNT FF data to Equation~\ref{eq:murnaghan}, which is computed as 
\begin{equation}
  {\rm{RMSE}} =  \left\lbrace \frac{1}{N} \sum_{i=1}^{N} \left[ P_{i}^{\rm MD} - P^{\rm Murn}(V_{i}^{\rm MD},K_0,K_0') \right]^2 \right\rbrace ^{1/2},
  \label{eq:murnaghanRMSD}
\end{equation}where the sum runs over the four $(P_{i},V_{i})$ state points sampled between 1~GPa and 10~GPa, inclusive. We limit our RMSE calculation to $P \leq 10$~GPa as this was the highest pressure sampled across all experiments; assessments at higher pressures are shown in the SI.

From the plot, it is clear that the error surface exhibits shallow curvature about the minimum and that there are many choices for $(K_0,K'_0)$ pairs that yield only modestly greater uncertainty. For instance, choosing the $(K_0,K'_0)$ fitted to the hydrostatic Stevens et al. data corresponds to a RMSE of 0.307~GPa and choosing $(K_0,K'_0)$ for the hydrostatic Bowden et al. data leads to an RMSE of slightly more than 1~GPa. Indeed, almost all of the experiments and all of the TNT FF versions yield $P(V)$ curves that differ by less than 1~GPa. While the above analysis indicates that the LJ version of the TNT FF is most accurate at high pressure, both the exp-6 and LJ FF versions are consistent with the spread in the available experiments for pressures up to $\approx$10~GPa. For this reason, we consider only the exp-6 TNT FF version in the remainder of this report.

\subsection{Melting Point} 
\label{subsec:melting-point}

Accurately capturing the melting point of TNT with an MD FF is important as the solid-liquid phase boundary can be crossed during static and dynamic loading events, and in the latter case can significantly influence hot spot formation processes\cite{Zhao2020,Kroonblawd2021,Das2021}. We characterized the accuracy of the melting point predicted by the TNT FF at atmospheric pressure through MD simulations of phase coexistence. We chose this approach because it: (1) is straightforward to implement;\cite{zhang_2012} (2) has been successfully applied to other organic HEs including TATB,\cite{mathew_2015, mathew_2018, Kroonblawd2024} RDX,\cite{Kroonblawd2022} and HMX;\cite{kroonblawd_2021_mechmat} and (3) because it has well-understood uncertainties and yields an upper bound on the melting point. It is empirically understood for related HEs that phase coexistence simulations predict the melting point to within 10\% of that obtained from much more complicated MD algorithms using thermodynamic integration.\cite{tow_2022}

\begin{figure}[t!]
  \centering
  \includegraphics[width=3.35in]{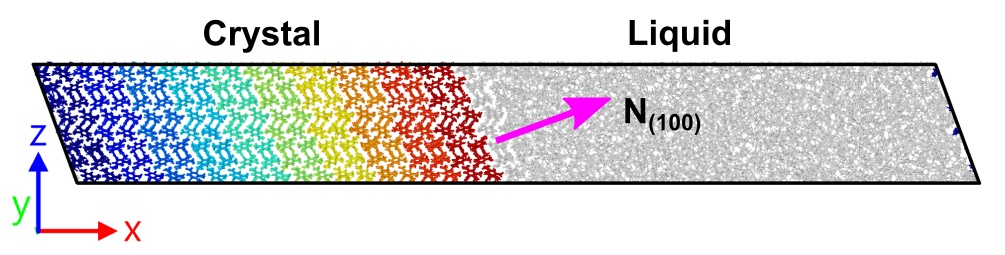}
  \caption{Phase coexistence simulation cell containing crystalline and liquid TNT.}
  \label{fig:twophase}
\end{figure}

The phase coexistence approach works by running NPT simulations with a simulation cell containing both crystal and liquid regions. When $T$ is set above the melting point $T_{\rm melt}$, the crystal dynamically melts resulting in an increase in state variables such as the internal energy $E$ as the thermostat supplies the latent heat of fusion. The simulation cell was built from a $20 \mathbf{a} \times 6 \mathbf{b} \times 2 \mathbf{c}$ replication of the TNT unit cell (1920 molecules) and had approximate $xyz$ dimensions $32~\rm{nm} \times 4~\rm{nm} \times 4~\rm{nm}$. The cell was divided along $x$ and half of the molecules were melted to prepare the initial configuration shown in Figure~\ref{fig:twophase} in which the liquid was in contact with (100) crystal facets. Independent 10~ns long NPT simulations were performed at specified temperatures set in 5~K increments over 325~K~$\leq T \leq 375$~K and the trajectories were inspected for signs of melting.

\begin{figure}[t!]
  \centering
  \includegraphics[width=3.35in]{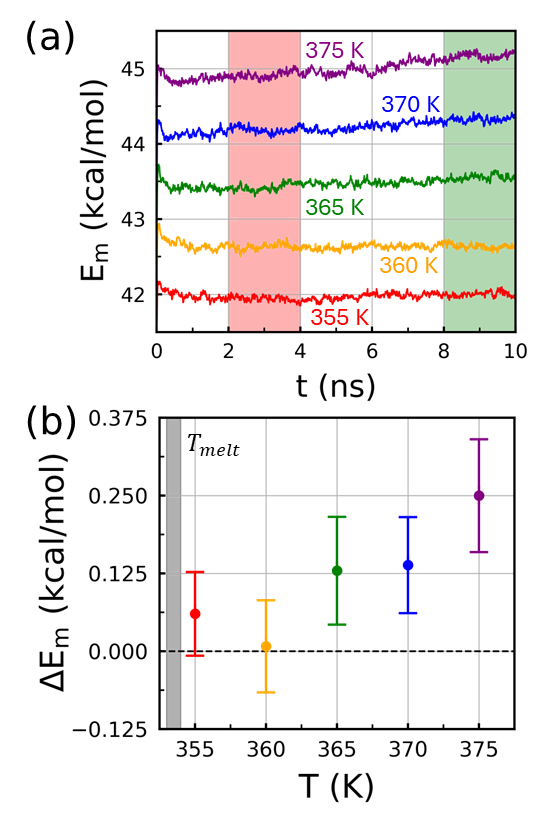}
  \caption{Analysis of phase coexistence simulations near the melting point showing (a) time histories for the molar internal energy $E_{\rm m}$ of simulations at constant selected temperatures, and (b) the time change in energy $\Delta E_{\rm m}$ between the two shaded regions in panel (a). The experimentally determined range for $T_{\rm melt}$ is shown as a grey bar in panel (b).} 
  \label{fig:melting-temp}
\end{figure}

Figure~\ref{fig:melting-temp}(a) shows resulting time histories of the molar internal energy $E_{\rm m}$ from NPT simulations near the melting point predicted by the TNT~FF. After a initial transient as the system equilibrates ($t < 1$~ns), several distinct features are apparent depending on the temperature. At 375~K, $E_{\rm m}$ steadily increases whereas in the 355~K case it remains approximately constant. These features are consistent with a first-order phase transition (i.e., melting) in the 375~K case and no phase change in the 355~K case. While the thermodynamic expectation is that $E_{\rm m}$ will decrease as the liquid crystallizes below $T_{\rm melt}$, crystallization of molecular systems is often a kinetically hindered process on MD timescales. For this reason, phase coexistence typically only yields a reliable upper bound on $T_{\rm melt}$ for molecular materials.

We inspected each energy time history for evidence of melting by determining whether a given temperature resulted in a statistically significant increase in $E_{\rm m}$ within the 10~ns timeframe of the simulations. The results from this analysis are shown in Figure~\ref{fig:melting-temp}(b), which plots the time change in energy
\begin{equation}
    \Delta E_{\rm m} = \bar{E}_{\rm m,8-10ns} - \bar{E}_{\rm m,2-4ns},
\label{eq:deltaE}
\end{equation}as a function of simulation temperature. Here, $\bar{E}_{\rm m}$ denotes an average energy determined over the subscript time intervals that are highlighted as shaded regions in panel (a). The standard deviation was taken as the uncertainty in $\bar{E}_{\rm m}$ and was propagated to obtain error bars on $\Delta E_{\rm m}$. From this analysis, we interpret the lowest temperature for which $\Delta E_{\rm m} > 0$ within uncertainty as an upper bound on $T_{\rm melt}$.

It is immediately clear from panel (b) that the TNT FF predicts a melting point at atmospheric pressure that is no greater than $T_{\rm melt} \leq 365$~K. This is within 4\% of the experimentally determined melting point, which is reported to be around 353-354~K.\cite{Vrcelj2003, Dattelbaum2014, Neyertz2012} Similar levels of accuracy were obtained for RDX and HMX using the Smith-Bharadwaj FF.\cite{tow_2022} Thus, we expect both the TNT~FF and our choice for the HMX~FF to accurately capture dynamic melting phenomena in simulations of the pure material and composite systems containing both materials.

\subsection{Thermal Response}
\label{subsec:isobars}

\begin{figure}[t!]
  \centering
  \includegraphics[width=3.35in]{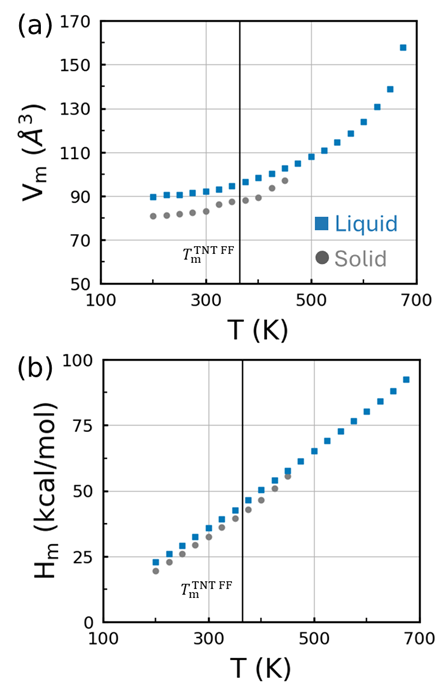}
  \caption{Isobaric temperature dependence of TNT crystal and liquid at atmospheric pressure, including (a) molar volume $V_{\rm m}$ and (b) molar enthalpy $\mathrm{H_{m}}$.}
  \label{fig:sl-isobar}
\end{figure}

We calculated the isobaric temperature response of TNT liquid and crystalline solid at atmospheric pressure through two additional series of NPT simulations. Simulations of the crystalline solid were performed starting from a $3 \mathbf{a} \times 6 \mathbf{b} \times 2 \mathbf{c}$ replication of the monoclinic unit cell and simulations of the liquid were performed using a cubic cell containing 400 molecules an in amorphous configuration. The molar volume $V_{\rm m}$ and enthalpy $H_{\rm m}$ were obtained from the last 250~ps of independent 500~ps long NPT simulations that were performed for temperatures in increments of 25~K over the interval 200~K~$\leq T \leq$ 700~K using both the solid and liquid starting cells. The results from these simulations are summarized in Figure~\ref{fig:sl-isobar} and thermodynamic properties derived from these simulations are discussed in the following subsection.

Focusing first on the solid phase, there are several distinct features in $V_{\rm m}$ as the temperature is increased, including above $T_{\rm melt}$ where the crystal is superheated yet metastable over the 500~ps long simulation. One of these is a subtle discontinuous jump below the predicted melting point between 300~K and 325~K. Inspection of the 300~K and 325~K equilibrium configurations (not shown) indicates that the structure retains the herringbone packing of the monoclinic phase, but with a distinct shift in the $\beta$ lattice angle as the temperature is increased. We also note that the packing of the 325~K structure is distinct from the metastable orthorhombic phase. This type of phase transition has been seen experimentally\cite{Dattelbaum2014}, although it has not been extensively characterized. There is another discontinuous jump above the melting point between 400~K and 425~K, which corresponds to a structural transition where lattice angles $\alpha \neq \gamma \neq 90^{\rm o}$. The crystal homogeneously melts within 500~ps above 450~K, which is nearly 100~K over the predicted $T_{\rm melt}$. Perhaps surprisingly, none of these structural transitions evident in $V_{\rm m}(T)$ or in the atomic configurations lead to discontinuous jumps or nonlinearity in $H_{\rm m}(T)$ for $T \leq 425~$K.

In comparison, the isobaric thermal response of the liquid/amorphous phase is more straightforward. The molar volume is smoothly varying but nonlinear over 200~K~$\leq T \leq 675$~K whereas the molar enthalpy depends linearly on temperature. Kinetics of crystallization are hindered on a 500~ps timescale, so the liquid becomes locked in a glassy solid phase below $T_{\rm melt}$. It is clear from both $V_{\rm m}(T)$ and $H_{\rm m}(T)$ that the glassy solid is distinct from the crystal and that it is enthalpically unfavorable. Our 700~K simulation predicted spontaneous vaporization of the liquid.

\subsection{Mechanical and Thermodynamic Properties}
\label{subsec:props}

A number of material properties can be derived from the information obtained in the preceding subsections using the TNT~FF, including the coefficient of thermal expansion (CTE) $\alpha_{V}$, the volume and enthlapy of fusion ($\Delta V_{\rm fus}$ and $\Delta H_{\rm fus}$), and the enthalpy of sublimation ($\Delta H_{\rm sub}$). Table~\ref{tab:TNT-props} summarizes this property data with comparisons to experiments given as validation testing. When known, we indicate whether a given property value was obtained for the monoclinic \(\mathrm{P2_1/a}\) phase or the orthorhombic \(\mathrm{Pca2_1}\) phase. Experimental entries for a given property are listed in chronological order in instances where multiple values are available for comparison. 

\begin{table}[t!]
\centering
\caption{Physical Properties of TNT}
\begin{tabular}{@{}llllr@{}}
\toprule
\multirow{1}{*}{Property}        
& \multirow{1}{*}{Value} & \multirow{1}{*}{Space Group} & \multirow{1}{*}{Temperature Interval (K)} & \multirow{1}{*}{Ref.}\\ 
\midrule
$\alpha_V^s < 300$ (K$^{-1}$)                              
& \(16.5 \times 10^{-5}\) & Unknown & 240-300 & \citenum{Heberlein1974}       \\ 
& \(4.95 \times 10^{-5}\) & Unknown & 233-300 & \citenum{Gibbs1980}       \\ 
& \(20.6 \times 10^{-5}\) & \(\mathrm{P2_1/a}\) & 123-300 & \citenum{Vrcelj2003}      \\
& \(19.8 \times 10^{-5}\) & \(\mathrm{Pca2_1}\) & 123-300 & \citenum{Vrcelj2003}      \\
& \(27.0 \times 10^{-5}\) & \(\mathrm{P2_1/a}\) & 200-300 & This work          \\
\midrule
$\alpha_V^s > 300$ (K$^{-1}$) 
& \(27.8 \times 10^{-5}\) & Unknown & 295-327 & \citenum{Eubank1950}      \\
& \(30.3 \times 10^{-5}\) & Unknown & 323-337 & \citenum{Rosen1959} \\
& \(5.34 \times 10^{-5}\) & Unknown & 300-333 & \citenum{Gibbs1980}       \\ 
& \(42.9 \times 10^{-5}\) & Unknown & 325-400 & This work          \\
\midrule
$\alpha_V^l$ (K$^{-1}$)                              
& \(87.2 \times 10^{-5}\) & --- & 356-393 & \citenum{Cady1962}  \\
& \(81.0 \times 10^{-5}\) & --- & 350-400 & This work          \\
 \midrule
$T_{\rm melt}$ (K)
& \(353-354\) & \(\mathrm{P2_1/a}\) & 353-354 & \citenum{Taylor1924,Rosen1959,Heberlein1974,Vrcelj2003}     \\ 
& \(365\)     & \(\mathrm{P2_1/a}\)           & 355-375 & This work   \\ 
 \midrule
 $\Delta V_{\rm fus}$ (A$^{3}$)
& 21.8   & \(\mathrm{P2_1/a}\) & 365 & This work     \\ 
 \midrule
\multirow{1}{*}{$\Delta H_{\rm fus}$ (kcal/mol)}
& \(4.88\)     & Unknown     & 353 & \citenum{tammann1913, urbanski1964chemistry}     \\ 
& \(5.4-5.6\)  & Unknown     & 353 & \citenum{Edwards1949}     \\ 
& \(5.34\)     & Unknown     & 354 & \citenum{Cady1962}     \\ 
& \(4.72\)     & \(\mathrm{P2_1/a}\)  & 353 & \citenum{Di2013}  \\
& 3.50     & \(\mathrm{P2_1/a}\) & 365 & This work     \\
\midrule
$\Delta H_{\rm sub}$ (kcal/mol)          
 & 28       & Unknown & 323-353 & \citenum{Edwards1949}     \\ 
 & 25.0     & Unknown & 327-349 & \citenum{Lenchitz1971}    \\
 & 23.7     & Unknown & 287-330 & \citenum{Pella1977}       \\
 & 26.5     & \(\mathrm{P2_1/a}\) & 1 & This work     \\
 & 23.9     & \(\mathrm{P2_1/a}\) & 300 & This work     \\
 \bottomrule\bottomrule
\end{tabular}
\label{tab:TNT-props}
\end{table}

Values for the CTE were determined from the volume-temperature data shown in Figure~\ref{fig:sl-isobar}. We consider separately the solid-phase CTE $\alpha_{V}^{s}$ below and above 300~K and the liquid-phase CTE $\alpha_{V}^{l}$. Values reported for the TNT FF in the table were calculated as
\begin{equation}
  \alpha_{V} = \frac{1}{V} \frac{\Delta V}{\Delta T}
  \label{eq:cte}
\end{equation}where the slope $\Delta V / \Delta T$ was determined through linear regression over the indicated temperature interval. Several experimental reports on the solid phase are available for comparison, but these comparisons are complicated by the solid-solid phase transition\cite{Heberlein1974, Vrcelj2003} and the fact that the majority of the reports\cite{Eubank1950, Rosen1959, Gibbs1980} did not characterize the crystal phase of the specimens studied. For instance, the experiments in Ref.~\citenum{Heberlein1974} started with the orthorhombic polymorph, but they observed a phase transition in the middle of the 100-300~K interval considered. In the table, we report the value they obtained after the phase transition, which is presumably a transition to the monoclinic phase. Vrcelj et al.\cite{Vrcelj2003} applied single-crystal x-ray diffraction to characterize their specimens. While they report that the monoclinic phase is more stable, they obtained CTEs for both phases up to room temperature and found the phases to be quite similar. An experimental value for the CTE of the liquid phase was obtained from a 1962 technical report by Cady and Howard.\cite{Cady1962}

The TNT~FF predicts a CTE for the solid phase that is similar to, if slightly larger than the available experiments. For example, the CTE predicted by the TNT~FF for the \(\mathrm{P2_1/a}\) phase is 31\% larger than than the value obtained by Vrcelj et al.\cite{Vrcelj2003} below 300~K. The most sigfnificant differences are seen in comparison against the values reported by Gibbs and Popolato,\cite{Gibbs1980} which are much smaller than both the TNT~FF values and the values obtained in other experimental studies. The TNT~FF also predicts a modestly larger CTE for the uncharacterized polymorph above 325~K, but the increase in CTE with increasing temperature is consistent with experiments. Both the TNT FF and experiment exhibit very similar CTEs for the liquid phase up to at least 400~K.

As discussed in Section~\ref{subsec:melting-point}, the TNT~FF predicts an upper bound on the melting temperature of $T_{\rm melt} = 365$~K, which is only 4\% larger than the narrow value range given across multiple experimental reports.\cite{Taylor1924, Rosen1959, Heberlein1974, Gibbs1980} We evaluated $\Delta V_{\rm fus}$ and $\Delta H_{\rm fus}$ at $T_{\rm melt} = 365$~K through linear interpolation of the data reported in Figure~\ref{fig:sl-isobar}. While we could not identify an experimental measurement of $\Delta V_{\rm fus}$, numerous experimental measurements of $\Delta H_{\rm fus}$ are available for comparison and indicate $ \Delta H_{\rm fus}$ is between 4.72 and 5.6 kcal/mol.\cite{tammann1913, urbanski1964chemistry, Edwards1949, Cady1962, Di2013} In comparison, the TNT FF predicts $\Delta H_{\rm fus} = 3.50$~kcal/mol, which is modestly smaller than experiment.

Finally, we consider the heat of sublimation $\Delta H_{\rm sub}$, which is a direct measure of the crystal binding energy. We computed $\Delta H_{\rm sub}$ from MD simulations as
\begin{equation}
    \Delta H_{\rm sub} =  \bar{E}_{\rm m, gas} - \bar{E}_{\rm m, crystal} + RT,
\label{eq:hsub}
\end{equation}where $\bar{E}_{\rm m, gas}$ and $\bar{E}_{\rm m, crystal}$ are the average potential molar energies of an isolated molecule and the crystal. Here, $\bar{E}_{\rm m, gas}$ was evaluated from the last half of a 100~ps long NVT trajectory and $\bar{E}_{\rm m, crystal}$ was obtained from the last half of a 100~ps long NPT trajectory performed at atmospheric pressure. As can be seen in the table, the TNT FF predicts that $\Delta H_{\rm sub}$ decreases with increasing temperature and is 23.9~kcal/mol at 300~K, which falls within the spread of experimental values.\cite{Edwards1949, Lenchitz1971, Pella1977}

Taken as a whole, the above assessments indicate that the TNT~FF reproduces the temperature-dependent equation of state of both solid and liguid TNT with good accuracy. Thermodynamic properties governing phase transitions indicate that the TNT FF reproduces the melting temperature to within 4\% (or 12~K), the latent heat of fusion is modestly underestimated by only $\approx$2~kcal/mol, and the latent heat of sublimation falls within experimental bounds. Thus, we expect that the TNT FF provides an accurate model for MD simulations involving the solid and liquid phases and transformations between these phases.

\section{TNT-HMX Composites}
\label{sec:composites}

\subsection{TNT-HMX FF Cross Interactions}
\label{sec:cross-terms}

As described in the Introduction, TNT is used in composite HE formulations where it is mixed with other energetic materials such as RDX and HMX. The previous section focused on calibrating and validating a TNT FF to assess its accuracy in MD simulations of pure TNT systems involving the solid and liquid phases. Here, we extend the TNT FF to model TNT-HMX systems by combining it with the established FF for HMX originally developed by Smith and Bharadwaj.\cite{Smith1999} The present subsection is focused on validating cross-interactions between the TNT and HMX FFs. Applications of the validated combined FF to model interface stability and melting dynamics of composite TNT-HMX systems are given in subsequent subsections.

Interactions between TNT and HMX molecules are computed using the exp-6-12 potential energy function with Coulomb electrostatics given in Equation~\ref{eq:HMX_bonded_nrg}. This function has three parameters $A$, $\rho$, and $C$ that must be specified for each unique pair of atom types in the TNT and HMX FFs, leading to $6 \times (5-1)/2 = 12$ sets of parameters $(A, \rho, C)$. The partial electric charges $q$ were not modified because doing so would alter the description of pure TNT and HMX. We consider here two common choices for mixing rules to compute the heteroatom $A_{\alpha\beta}$, $\rho_{\alpha\beta}$, and $C_{\alpha\beta}$ from the homoatom values $A_{\alpha\alpha}$, $\rho_{\alpha\alpha}$, and $C_{\alpha\alpha}$ listed in Tables~\ref{tab:HMX-nb} and \ref{tab:TNT-nb}.
The first choice, which we denote ``Mix 1,'' combines $A$, $\rho$, and $C$ using standard Lorentz-Berthelot mixing rules,\cite{Mayo1990, White2000} which adopt geometric means for $A$ and $C$ and an arithmetic mean for $\rho$. The second choice, denoted ``Mix 2,'' applies geometric means to $A$, $\rho$, and $C$. This second choice is the same one originally proposed for HMX-HMX nonbonded interactions in the HMX~FF\cite{Smith1999} and that we adopted for TNT-TNT nonbonded interactions in the TNT~FF.

\begin{figure*}[t!]
  \centering
  \includegraphics[width=0.98\textwidth]{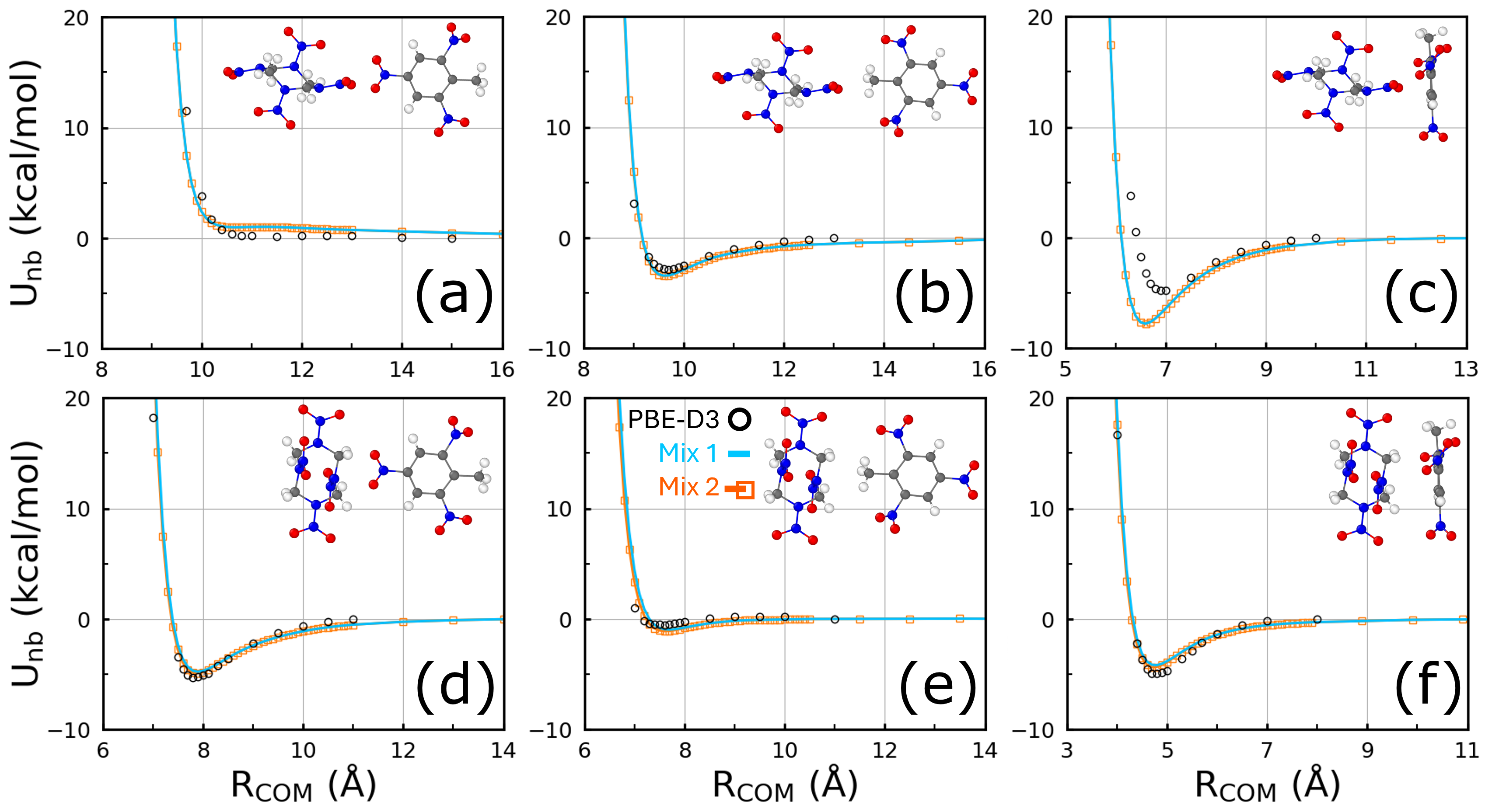}
  \caption{Nonbonded potential energy surfaces $U_{\rm nb}$ obtained for oriented TNT-HMX dimers using PBE-D3 DFT calculations and two choices for TNT-HMX FF mixing rules. The dimer orientations respectively emphasize interactions and close contacts between specific functional groups, including (a) HMX nitro $\rightarrow$ TNT nitro, (b) HMX nitro $\rightarrow$ TNT methyl, (c) HMX nitro $\rightarrow$ TNT aromatic ring, (d) HMX ring $\rightarrow$ TNT nitro, (e) HMX ring $\rightarrow$ TNT methyl, and (f) HMX ring $\rightarrow$ TNT aromatic ring.}
  \label{fig:cross-scans}
\end{figure*}

Accuracy of a given mixing rule choice was assessed through comparison to DFT calculations of potential energy surfaces for the six oriented TNT-HMX dimer pairs shown in Figure~\ref{fig:cross-scans}. All potential energy surfaces were computed as a series of single-point energy evaluations for rigid-body displacements $R_{\rm COM}$ along the axis separating the molecular centers of mass. Treating the molecules as rigid bodies effectively isolates the nonbonded potential energy $U_{\rm{nb}}$. Rotational orientations of the two molecules were determined by first aligning the principal rotational axes of each molecule with a Cartesian lab frame and then rotating each molecule to test close contacts between specific functional groups. For instance, the dimer pair in panel (a) was oriented to sample close contacts between nitro groups on TNT and HMX.

Inspection of each panel in Figure \ref{fig:cross-scans} shows that both mixing rule choices yield very similar results, but Mix 2 is consistently (if marginally) closer to the DFT PBE-D3 baseline. Interactions for dimer pair (a) are purely repulsive and both DFT and the FF exhibit strong repulsion at very similar displacements. The other five dimer pairs all exhibit potential energy surfaces with a local minimum. Most of these pairs exhibit both well depths and minimum energy separation distances that are exceedingly close to DFT. The FF predictions for dimer pair (c) are the only ones that exhibit a notable deviation from DFT; here the well depth is overpredicted by 2.5~kcal/mol and the minimum energy separation distance occurs at 6.5~{\AA} rather than at 7.0~{\AA}. The fact that discrepancies are greatest for dimer (c) is perhaps not unexpected as this dimer probes interactions that are governed by delocalized electrons in both the HMX nitro group and TNT aromatic ring.

Overall, the accuracy of the FF predictions for TNT-HMX interactions is excellent, with deviations that are generally smaller than those obtained for dimer interactions in other \textit{pure} HE materials\cite{Kroonblawd2024} and between HEs and non-HEs.\cite{scher_2024_chemrxiv} While we did not directly test TNT-RDX pairs, we anticipate similar levels of accuracy in the application of the combined FF to TNT-RDX systems. This is in part due to the close chemical similarity of RDX and HMX and the fact that the Smith-Bharadwaj FF uses the same set of parameters to describe both RDX and HMX. In future work, explicitly fitting the TNT-HMX cross interactions to the DFT data in Figure~\ref{fig:cross-scans} could further improve accuracy,\cite{scher_2024_chemrxiv} especially for dimer pair (c).

\subsection{Surface and Interface Energies}

Energetic stability of crystal facets influences the morphology of grown crystals and can also induce anisotropy in surface phenomena. Generally, crystal facets with lower surface energy in a given solvent are expressed with greater proportion in precipitated crystals.\cite{Brahmbhatt2024, li_2024} The impact of surface energy on surface phenomena can be nuanced. In the context of surface-initiated melting of molecular HE materials, areal packing density and molecular bonding can strongly impact melting kinetics and the propensity for superheating on nanosecond timescales.\cite{mathew_2015} While the energetic stability of interfaces in HE materials is relatively unexplored, the interface energy can help explain wetting behavior and guide identification of favorable material interactions.\cite{gee_2007} For this reason, we anticipate that MD predictions for surface and interface energies of TNT-HMX systems can help explain a variety of phenomena, including the kinetics of melting and crystallization, which is of particular relevance for melt-castable HE formulations.

\begin{table}[t!]
\centering
\caption{Equilibrium Lattice Parameters for TNT and HMX at 10~K and 1 atm}
\begin{tabular}{@{}lcccccc@{}}
\toprule
\textbf{Material} & $a$ (\AA) & $b$ (\AA) & $c$ (\AA) & $\alpha$ (deg.) & $\beta$ (deg.) & $\gamma$ (deg.) \\ 
\midrule
TNT & 15.134 & 6.2414 & 19.785 & 90.000 & 113.15 & 90.000 \\
HMX & 6.5489 & 10.264 & 7.6077 & 90.000 & 98.430 & 90.000 \\ 
\bottomrule\bottomrule
\end{tabular}
\label{table:lattice_parameters}
\end{table}

We focus here on surfaces and interfaces between two distinct TNT crystal facets, namely (010) and (001), and the lowest-energy\cite{Brahmbhatt2024} and most expressed\cite{gallagher_2014} HMX crystal facet, namely (011). All systems with free surfaces and crystal-crystal interfaces were prepared using the GCCM code\cite{Kroonblawd2016} with equilibrium lattice parameters determined for both crystals at 10~K and 1~atm using NPT simulations (see Table~\ref{table:lattice_parameters}). By convention, all calculations used 3D periodic boundaries with cells that were oriented so that the selected surface (or interface) normal vectors were aligned with the $z$ direction in a Cartesian lab frame.

The GCCM code identifies possible crystal-crystal interfaces between two facets that have lattice mismatches within prescribed strain tolerances. Solutions for commensurate interfaces for joining two crystal grains A and B are given by two sets of cell parameters $(\mathbf{x}^{\rm{A}}_{1},\mathbf{x}^{\rm{A}}_{2},\mathbf{x}^{\rm{A}}_{3})$ and $(\mathbf{x}^{\rm{B}}_{1},\mathbf{x}^{\rm{B}}_{2},\mathbf{x}^{\rm{B}}_{3})$ where the $\mathbf{x}_i$ are lattice parameters. That is,
\begin{equation}
    \mathbf{x}_i = m_i \mathbf{a} + n_i \mathbf{b} + p_i \mathbf{c},
    \label{eq:gccmvec}
\end{equation}where $m_i$, $n_i$, and $p_i$ are integers and $\mathbf{a}$, $\mathbf{b}$, and $\mathbf{c}$ are the primitive lattice vectors for the crystal. The two crystal grains are joined along their respective $\mathbf{x}_1$-$\mathbf{x}_2$ planes with a total surface area given by 
\begin{equation}
    A = \mathbf{x}_1 \cdot \mathbf{x}_2.
    \label{eq:x-area}
\end{equation}For the purpose of solution selection, the area examined is a max side length area calculated as 
\begin{equation}
    A_{\rm msl} = (\max\{\mathbf{x}_1, \mathbf{x}_2\})^2
    \label{eq:max-area}
\end{equation}This ensures that systems with similar side lengths are given preference. Both grains must be placed in a single simulation cell with cell vectors $(\mathbf{x}_{1},\mathbf{x}_{2},\mathbf{x}_{3})$ subject to the following constraints,
\begin{equation}
    \mathbf{x}_1 = \frac{\mathbf{x}^{\rm{A}}_{1} + \mathbf{x}^{\rm{B}}_{1} }{2}, 
    \label{eq:gccmConstrainX1}
\end{equation}and
\begin{equation}
    \mathbf{x}_2 = \frac{\mathbf{x}^{\rm{A}}_{2} + \mathbf{x}^{\rm{B}}_{2} }{2}, 
    \label{eq:gccmConstrainX2}
\end{equation}and 
\begin{equation}
    \hat{\mathbf{x}}^{\rm{A}}_{3} = \hat{\mathbf{x}}^{\rm{B}}_{3}  
    \label{eq:gccmConstrainX3}
\end{equation}which in most cases requires straining the individual grains. Good GCCM solutions balance the system size (e.g., $A$ or $A_{\rm msl}$) and the strain, which we measure as 
\begin{equation}
    \epsilon =
    \left(\frac{ x^{\rm{A}}_{1} - x^{\rm{B}}_{1} }{ x^{\rm{A}}_{1} }\right)^2    
    + \left(\frac{ x^{\rm{A}}_{2} - x^{\rm{B}}_{2} }{ x^{\rm{A}}_{2} }\right)^2 
    + (\theta_{\rm{A}} - \theta_{\rm{B}} )^2 
    + (\phi_{\rm{A}} - \phi_{\rm{B}} )^2
    + (\psi_{\rm{A}} - \psi_{\rm{B}} )^2,
    \label{eq:strain}
\end{equation}where $x_i = | \mathbf{x}_i |$ and the angles are defined as $\theta \angle \mathbf{x}_1 \mathbf{x}_2$, $\phi \angle \mathbf{x}_1 \mathbf{x}_3$, and $\psi \angle \mathbf{x}_2 \mathbf{x}_3$. We chose GCCM solutions for constructing interfaces that had low strain $\epsilon$ while maintaining approximately square transverse dimensions (i.e., $x_1 \approx x_2$), similar sizes across all surface and interface systems, and that had a total area $A \leq 7500$~{\AA}$^{2}$. Additional discussion on obtaining optimum GCCM solutions is given in the SI.

Surface energies were determined using crystal slabs with vacuum space added along the $z$ direction, which leads to two free surfaces given the 3D periodic boundary. The surface energy was calculated as 
\begin{equation}
  \gamma = \frac{E_{\rm{slab}} - N_{\rm{slab}} E_{\rm{m}}^{\rm{bulk}} }{2A},
  \label{eq:surfaceEnergy}
\end{equation}where $E^{\rm{slab}}$ is the NVT-average energy of the crystal slab with exposed vacuum, $N_{\rm{slab}}$ is the number of molecules in the crystal slab, $E_{\rm{m}}^{\rm{bulk}}$ is the average energy per molecule of the 3D periodic bulk crystal, and $A$ is the area of the simulation cell in the $xy$ plane. The factor of two arises as the slab simulation has two free surfaces. Transverse dimensions of all slab cells were chosen to be at least 3$\times$ larger than the FF cutoff (11~{\AA}) and the slab thickness along $z$ was set to $20.0 \pm 0.4$~nm. All energies were obtained at 10~K from the last half of a 250~ps long NVT simulation.

Interface energies were determined using simulation cells containing two oriented crystal slabs that were stacked along the $z$ direction, which leads to two crystal-crystal interfaces given the 3D periodic boundary. The interface energy was calculated as 
\begin{equation}
  \Gamma = \frac{ E_{\rm{int}} - \left( N_{\rm{slab,1}} E_{\rm{m}}^{\rm{bulk,1}} + N_{\rm{slab,2}} E_{\rm{m}}^{\rm{bulk,2}} \right) }{2A},
  \label{eq:interfaceEnergy}
\end{equation}where $E_{\rm{int}}$ is the average energy of the interface system, $N_{\rm{slab,1}}$ and $N_{\rm{slab,2}}$ are the numbers of molecules in crystal slabs 1 and 2, $E_{\rm{m}}^{\rm{bulk,1}}$ and $E_{\rm{m}}^{\rm{bulk,2}}$ are the average energies per molecule of the 3D periodic bulk crystals 1 and 2, and $A$ is the area of the simulation cell in the $xy$ plane. The above expression for $\Gamma$ measures the favorability of forming an interface relative to the infinite bulk. Favorability of an interface forming between crystals with pre-exposed surfaces is therefore  
\begin{equation}
  \Gamma^{'} = \Gamma - (\gamma_1 + \gamma_2),
  \label{eq:interfaceEnergyAlt}
\end{equation}where $\gamma_1$ and $\gamma_2$ are the surface energies for the two crystal slabs computed using Equation~\ref{eq:surfaceEnergy}. The transverse dimensions of the interface cells were chosen to be at least 3$\times$ larger than the FF cutoff (11~{\AA}) and each crystal slab was $20.0 \pm 0.4$~nm thick along the $z$ direction. Unlike with the surface cells, the strains imposed to match the crystal slabs at an interface can induce stresses in the system. We obtained $E_{\rm{int}}$ from a two-stage trajectory in which we first ran 250~ps of NVT dynamics followed by 250~ps of NPT dynamics with the thermostat set to 10~K and the barostat set to 1~atm. The average value for $E_{\rm{int}}$ was determined from the last half of the NPT portion of the trajectory.

\begin{table}[t!]
\centering
\caption{Surface and interface energies for TNT and HMX at 10~K and 1 atm}
\begin{tabular}{@{}lc@{}}
\hline \hline 
Property & Energy (mJ/m$^2$) \\ 
\hline
$\gamma_{\rm{TNT(010)}}$ & $142 \pm 5$ \\
$\gamma_{\rm{TNT(001)}}$ & $109 \pm 5$ \\
$\gamma_{\rm{HMX(011)}}$ & $177 \pm 5$ \\
\hline
$\Gamma_{\rm{TNT(010)|HMX(011)}}$ & $250 \pm 5$ \\
$\Gamma_{\rm{TNT(001)|HMX(011)}}$ & $201 \pm 5$ \\
\hline
$\Gamma^{'}_{\rm{TNT(010)|HMX(011)}}$ & $-69 \pm 9$ \\
$\Gamma^{'}_{\rm{TNT(001)|HMX(011)}}$ & $-85 \pm 9$ \\
\hline \hline
\end{tabular}
  \label{tab:SE-table}
\end{table}

Table~\ref{tab:SE-table} collects the resulting surface and interface energies for our selected system configurations. Given the above definitions for $\gamma$, $\Gamma$, and $\Gamma^{'}$, a positive value indicates that the formation of a surface/interface is energetically unfavorable whereas a negative value indicates that the formation is favorable. Both TNT crystal facets exhibit positive-valued surface energies that are lower than than the (011) HMX facet, indicating that both TNT facets are more stable than the HMX one. It is also clear that the (001) TNT facet is more stable than the (010) facet. The computed interface energies indicate that HMX interfaces with the high-energy (010) TNT facet are less favorable than with the low-energy (001) TNT facet. Positive $\Gamma$ indicates that interfaces are not energetically favorable and that there may be a thermodynamic driver for reducing interface area between TNT and HMX. At the same time, negative $\Gamma^{'}$ indicates that adhesion between TNT and HMX is favorable if free crystal surfaces are already present. The above data indicate that there is an energetic driver for reducing grain surface area in TNT-HMX composites, which would promote grain coarsening. However, predicting the evolution of TNT-HMX composite microstructure involving both interfaces and voids is likely to be nuanced and would also depend on kinetic barriers that are not assessed by these simple measures.

\subsection{Microstructural Effects on TNT Melting Dynamics}

Localized melting and amorphization can impact heat generation during hot spot formation processes\cite{Zhao2020, Kroonblawd2021, Das2021} and can also accelerate chemical reaction kinetics in non-melt-castable HEs such as RDX and TATB.\cite{Sakano2018, Kroonblawd2020, hamilton_2022_jpcl, hamilton_2023} However, liquid TNT is chemically stable, so endothermic melting processes might serve to reduce energy localization into hot spots and enhance safety of TNT-containing HE formations. While it is understood that microstructural features such as voids and interfaces can serve as nucleation points for melting, the influence of these defects on melting dynamics is not well understood in molecular HE crystals. To this end, we apply our TNT-HMX FF to perform MD simulations of dynamic melting of TNT in a variety of plausible microstructural environments.

Melting of TNT in five distinct microstructural environments was considered, including perfect bulk TNT crystal without defects, surface-initiated melting in TNT with exposed (010) and (001) surfaces, and melting in idealized TNT-HMX composites at interfaces between the (010) and (001) TNT crystal facets and the (011) HMX facet. Each simulation was constructed so as to contain a similar quantity of TNT crystal and was heated from 10~K to 700~K with the temperature incremented by 10~K every 100~ps. All simulations were performed with NPT dynamics. Surface simulations had three barostats applied to the transverse dimensions and simulations of bulk and interface systems were performed using NPT dynamics with a triclinic barostat set to 1~atm. Five separate simulations were performed for each of the five cases starting from different initial atomic velocities sampled at 10~K.

\begin{figure}[t!]
  \centering
  \includegraphics[width=0.98\textwidth]{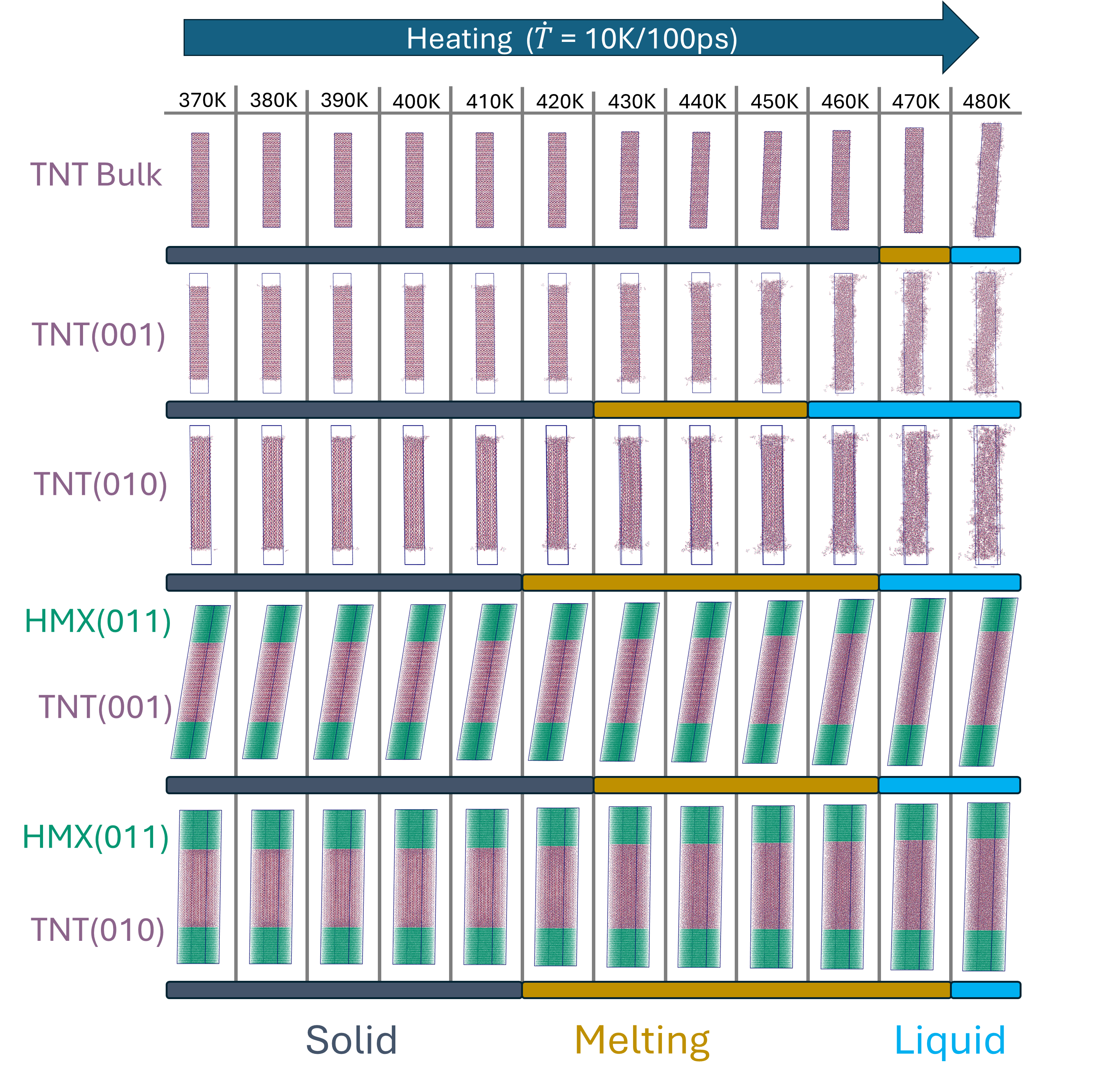}
  \caption{Series of configuration snapshots showing example simulations of dynamic melting of TNT in each of the five different microstructural environment cases. TNT and HMX molecules are respectively colored purple and green and molecules are rendered in unwrapped coordinates to better highlight disordering and diffusion. The primary simulation cell is rendered with black lines. Portions of each trajectory are labeled ``solid,'' ``melting,'' and ``liquid'' based on an analysis of thermodynamic and structural information averaged across five independent simulations performed for each case.}
  \label{fig:vis-melting}
\end{figure}

We focus first on the qualitative characteristics of TNT melting during dynamic heating. Figure~\ref{fig:vis-melting} shows example configuration snapshots of the time/temperature history for each of the five cases. Numerical quantification of the onset and progress of TNT melting based on thermodynamic and structural information is discussed in more detail below. Note that the figure snapshots show temperature states that are between the FF-predicted melting points for pure TNT and HMX, which are respectively 365~K and 579~K.\cite{tow_2022} Inspection shows that without defects, TNT can be superheated by as much as 100~K on a 1-ns timescale before it rapidly and homogeneously melts. In contrast, both the surface and interface systems exhibit more modest superheating. Melting in both the surface and interface systems starts from the defect site and proceeds in a sequential layer-by-layer fashion into the crystal slab. Visual inspection of the snapshots also indicates that TNT melting might proceed more quickly from (010) facets than from (001) facets, which tracks with heuristic expectations based on the computed surface energies in Table~\ref{tab:SE-table}. That is, one might expect slower melting normal to the energetically more-stable (001) facet.

To better quantify TNT melting dynamics, we assessed changes in orientational order through the second Legendre polynomial of a molecular orientation angle $\phi$ as 
\begin{equation}
  P_{2}[ \cos(\phi) ] = \frac{1}{2} \left[ 3\cos^2(\phi) - 1 \right].
  \label{eq:p2}
\end{equation}Here, $\phi$ for a TNT molecule is defined at a point in time $t$ with reference to an earlier time $t_0$ as
\begin{equation}
  \cos(\phi) = \mathbf{Q}(t) \cdot \mathbf{Q}(t_0),
  \label{eq:p2-q}
\end{equation}where $\mathbf{Q}(t)$ is the unit normal vector for the TNT C$_6$ ring at time $t$. 
We computed $\mathbf{Q}(t)$ by identifying a best-fit plane for all 6 ring atoms through least squares fitting. 
All orientation angles were measured relative to the orientation of TNT molecules at the end of the 300~K portion of each thermal ramp trajectory. The $P_{2}$ order parameter is bounded between -0.5 and 1.0, with respective values of -0.5, 0.0, and 1.0 when $\mathbf{Q}(t)$ is antiparallel, orthogonal, and parallel to the reference orientation. Thus defined, an average $P_{2}$ value of 0.0 indicates complete loss of the original reference order.

\begin{figure}[t!]
  \centering
  \includegraphics[width=0.98\textwidth]{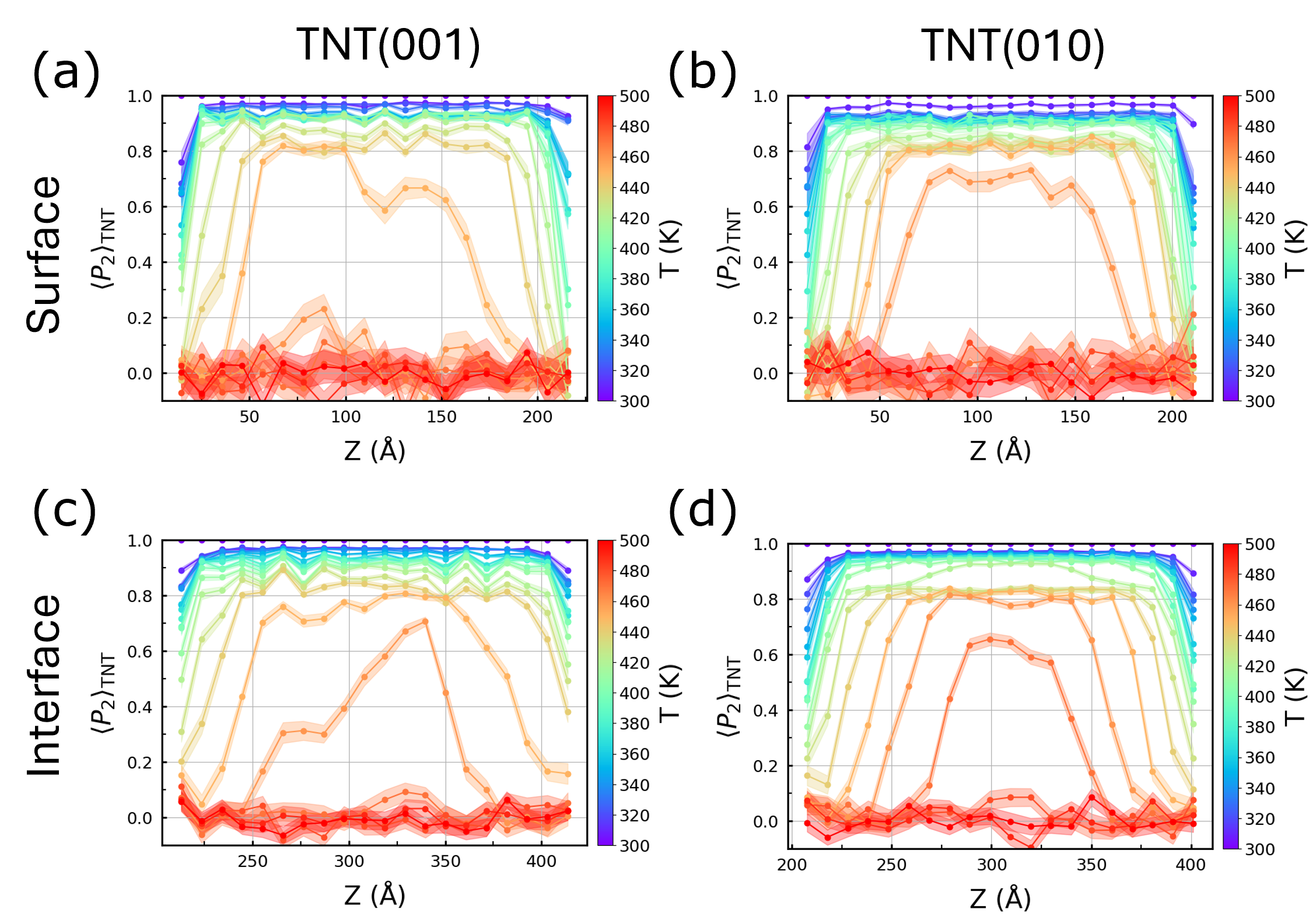}
  \caption{Spatially resolved $P_2$ order parameter of TNT molecules plotted as a function of position $z$ for the four cases involving free surfaces or interfaces. Each panel shows $P_2$ at selected time/temperature points during the MD trajectory, with curve colors denoting the temperature value. Panels are organized to show the low-energy (001) facet results on the left and high-energy (010) facet results on the right, with the top row corresponding to free surface systems and the bottom row to TNT-HMX interface systems. Error bars correspond to the standard error computed over all molecules within a given Lagrangian bin.}
  \label{fig:OP-matrix}
\end{figure}

Figure~\ref{fig:OP-matrix} shows the spatially resolved average $\left\langle P_{2} \right\rangle_{\rm{TNT}}$ order parameter for the surface and interface systems at selected times/temperatures. Values for $\left\langle P_{2} \right\rangle_{\rm{TNT}}$ were obtained by Lagrangian binning normal to the surface/interface along $z$ and averages were taken over all molecules within a bin. Each panel corresponds to a single representative trajectory chosen from the ensemble of five independent trajectories. (An analogous plot for the bulk system is provided in the SI.) Each case clearly exhibits a layer-by-layer melting process that initiates from both free surfaces/interfaces and progresses in an approximately symmetric manner towards the center of the slab.

\begin{figure}[t!]
  \centering
  \includegraphics[width=0.49\textwidth]{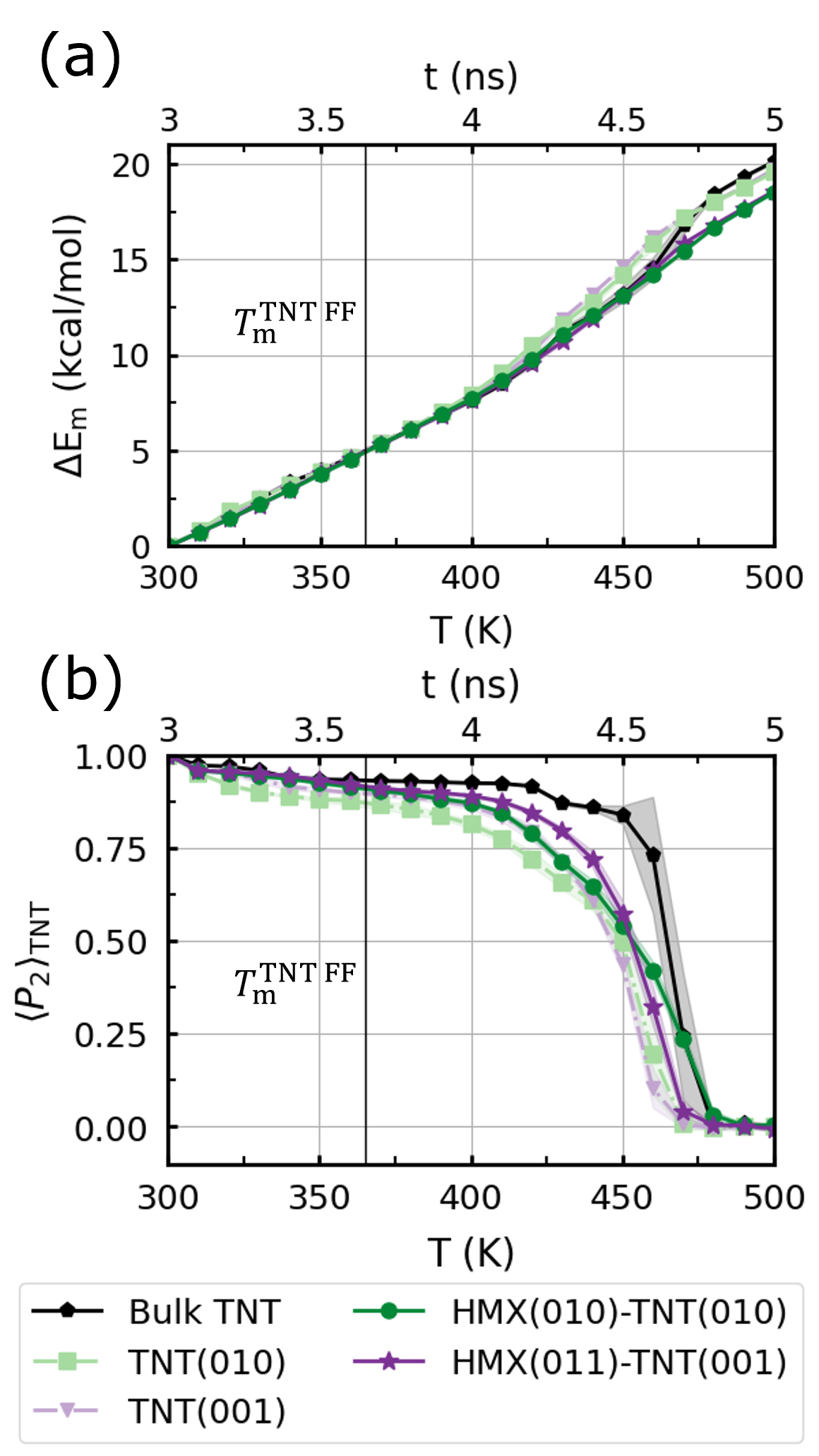}
  \caption{(a) Change in molar internal energy of the TNT molecules measured with respect to $E_{\rm m}(T = 300~\rm{K})$. (b) Evolution of the system-average $P_{2}$ order parameter of the TNT molecular orientations. Symbols correspond to the mean and shaded regions about each curve to the standard deviation obtained from five independent simulations for each case. Both plots are shown on dual $x$-axes of temperature (bottom axis) and simulation time (top axis).}
  \label{fig:quant-melting}
\end{figure}

System-level ensemble averages of the change molar internal energy of the TNT molecules $\Delta E_{\rm m}$ and of the $P_{2}$ order parameter were computed for all five cases to more clearly identify the onset, progression, and completion of melting. These data are shown in Figure~\ref{fig:quant-melting}. Here, we compute and plot the ensemble mean and standard deviation by averaging over all five independent simulations for each case. Note that $\Delta E_{\rm m}$ measures only the energy of TNT molecules, including in systems containing HMX.

Focusing first on $\Delta E_{\rm m}$ in panel (a), it is clear that the bulk, surface, and interface systems each exhibit different trends with heating and dynamic melting. Each system follows the same linear increase with temperature up to $\approx$400~K. Departure from this linear trend coincides with the onset melting as the thermostat supplies the latent heat of fusion. However, the trends in $\Delta E_{\rm m}$ are subtle to differentiate between the various cases. We note that the TNT liquid exhibits modest differences in $\Delta E_{\rm m}$, which can be seen at 500~K. Compared to the bulk system, both surface systems exhibit a marginally smaller value and the interface systems exhibit a $\Delta E_{\rm m}$ that is 2~kcal/mol smaller than the bulk. These differences may be partially due to a kinetic effect (e.g., melting is incomplete), but might also indicate that the TNT liquid in both the surface and interface cases exhibits non-bulk-like thermodynamic properties.

The system-average $P_{2}$ order parameter shows much clearer trends with melting and also clearer differences between the five different cases. Among the five cases, bulk TNT is most prone to superheating and also exhibits the least variance in the melting nucleation time/temperature. All five bulk simulations stay in the ordered solid phase up to 450~K and a few stay in the solid phase up to 460~K, which is reflected in the large error bars for the 460~K temperature point. All of the bulk simulations completely melt within a time/temperature interval of 200~ps/20~K. The rate of change in $P_{2}$ is also greatest for the bulk simulations.

In contrast to the bulk, the surface and interface systems exhibit slower melting processes that occur over a longer time/temperature interval. Surface-initiated melting occurs earlier on average for the high-energy (010) facet as compared to the low-energy (001) facet, but melting from the (010) facet also progresses more slowly. A similar trend is seen for the interface systems, wherein melting initiates earlier in the (010) system and finishes melting after the (001) system is fully melted. While both free surfaces and crystal-crystal interfaces provide effective nucleation points for melting, on average, surface-initiated melting starts and finishes more quickly than interface-initiated melting for both crystal facets.

The above result that the (010) cases initiate melting earlier but take longer overall to fully melt is plausibly due to the geometry of TNT crystal packing. The average TNT layer separation distance is 3.0~{\AA} for stacking normal to the (010) facet whereas this distance is 4.9~{\AA} for stacking normal the (001) facet. This means that for a fixed slab thickness ($\approx$20~nm), there are 60\% more distinct TNT crystal layers that must be melted in the (010) systems as compared to the (001) systems. The fact that melting initiates earlier for the high-energy (010) facet as compared to the low-energy (001) facet tracks with physical expectations based on the surface and interface energies. That is, one would expect high-energy surfaces/interfaces to be less energetically stable and therefore better nucleation points for melting. However, our simulations indicate that the overall rate of surface/interface initiated melting, for instance measured in terms of a melt front velocity, is not determined by the surface energy alone.

\section{Conclusions}

Formulated HEs are indispensable composite materials that impact many technological areas including construction/demolition, mining, defense, and propulsion. However, predicting the initiation safety properties of HEs from basic material properties is complicated by the fact that initiation is governed by hot spots that form at material microstructural defects. While hot spot formation is increasingly well understood in pure materials, very little is known about fundamental hot spot formation processes in multi-material composites, which is a materials class that encompasses most HEs used in practice. Dynamical, micron-scale hot spot formation processes remain inaccessible to experimental measurements, so direct numerical simulations are likely to be the main route to characterize and understand these processes at microstructural defects for the foreseeable future.

To this end, we developed all-atom MD FF models for large-scale simulations of real HE formulations based on TNT and the nitramine HEs HMX and RDX. The new TNT FF presented here was calibrated to data on the molecular vibrational spectrum and high-pressure equation of state and was shown to accurately capture the thermodynamics of TNT under thermal and pressure loading, including phase transitions. A combined FF to model TNT-HMX systems was validated against DFT calculations of the intermolecular potential energy surface and exhibits excellent accuracy that is equivalent or superior to established FFs for pure HE materials.

We applied our new TNT-HMX FF to predict the energetic stability of microstructural defects, including free surfaces and TNT-HMX crystal-crystal interfaces. These calculations indicate that there is an energetic driver for grain coarsening in TNT-HMX composites. Simulations of TNT melting in plausible microstructural environments under a dynamically evolving thermal load were performed to better understand the coupling between microstructure and phase transitions in composite HEs. Our simulations show that free surfaces and TNT-HMX interfaces are both effective nucleation points for melting. We also uncover evidence for correlations between the energetic stability of a given defect and the propensity for the TNT crystal to superheat. However, our simulations show that superheating by 50-100~K is to be expected on sub-ns timescales, even with defects to nucleate melting. These timescales for superheating are similar to timescales for shock-induced hot spot formation. Determining dynamic loading regimes where endothermic melting processes can effectively reduce peak hot spot temperatures in TNT-based HEs is likely to be complicated, but these regimes can be predicted using direct numerical simulations with the TNT-HMX FF.

The HE FFs and modeling framework presented here are anticipated to enable a wide range of future studies on hot spot formation in composite HEs. These efficient and accurate MD models, coupled with modern GPU-accelerated computers, are suitable for performing direct numerical simulations of hot spot formation on spatial scales reaching several square microns.\cite{Kroonblawd2025} Such MD simulations will be crucial for resolving complicated material dynamics that is governed by strong coupling between microstructural defects and interfaces, material mechanics, momentum and energy transport, phase transitions, and chemistry. Developing a better physical understanding of the role of interfaces in hot spot formation will help improve computational models to predict HE initiation in intentional-use and accident scenarios and holds promise for accelerating the design and qualification of new HE formulations with enhanced properties.

\begin{acknowledgement}

This work was performed under the auspices of the U.S. Department of Energy by Lawrence Livermore National Laboratory under Contract DE-AC52-07NA27344. Work by Purdue University was supported by LLNL subcontract B648789. Approved for unlimited release, LLNL-JRNL-2003615-DRAFT. Approved for unlimited release, LA-UR-25-22597.

The authors acknowledge RCAC and LLNL for computing resources. MPK thanks Tommy Sewell for useful discussions regarding the HMX force field. BWH contributed to this work while at Purdue University.

\end{acknowledgement}


\begin{suppinfo}

Supplemental Information below

\end{suppinfo}


\bibliography{bibtex}

\end{document}